\documentclass[12pt]{article}

\usepackage[T2A]{fontenc}
\usepackage[intlimits]{amsmath}
\usepackage{amsmath}
\usepackage{epsfig}
\usepackage{cite}
\usepackage{epsf}
\usepackage{graphicx}
\usepackage{epstopdf}
\usepackage{cmap}
\usepackage[labelsep=period]{caption}

\newcommand{\be}{\begin{equation}}
\newcommand{\ee}{\end{equation}}
\newcommand{\ben}{\begin{equation*}}
\newcommand{\een}{\end{equation*}}

\newcommand{\notp}[2][]{#2\!\!\!/_{#1\bot}}

\newcommand{\Tr}{\mathrm{Tr}}
\renewcommand{\not}[1]{#1\mspace{-8mu}\slash}

\newcommand{\q}{\vec{q}}
\newcommand{\qs}{\vec{q}^{\;2}}

\newcommand{\x}{\vec{r}}
\newcommand{\xs}{\vec{r}^{\;2}}
\newcommand{\xp}{\vec{r}^{\;\prime}}

\begin{document}
\numberwithin{equation}{section}

\begin{titlepage}
\hskip 11cm \vbox{ \hbox{Budker INP 2014-19}  }
\vskip 3cm
\begin{center}
{\bf  Gluon Reggeization  in Yang-Mills Theories $^{~\ast}$}

\end{center}
\vskip 0.5cm \centerline{ V.~S.~Fadin$^{\dag}$, M.~G.~Kozlov$^{\dag\dag}$,
A.~V.~Reznichenko$^{\ddag}$ } \vskip .3cm
\begin{center}
{\sl Budker Institute of  Nuclear Physics of Siberian Branch Russian Academy of Sciences,  Novosibirsk, 630090 Russia,\\
 Novosibirsk State University, Novosibirsk, 630090 Russia}
\end{center}
\vskip 1cm

\begin{abstract}
The proof of the multi-Regge form of multiple production amplitudes in  the
next-to-leading logarithmic approximation is presented for   Yang-Mills  theories
with  fermions and scalars in any representations   of the colour group and  with
any Yukawa-type interaction.  Explicit  expressions for the  Reggeized gauge boson
trajectory, the  Reggeon vertices and the impact factors are given. Fulfilment of
the  bootstrap conditions is proved.

\end{abstract}

\hrule \vskip .3cm \noindent $^{\ast}${\it  Work is supported  by the Russian
Scientific Foundation, grant} RFBR grant 13-02-01023.

\vfill $
\begin{array}{ll} ^{\dag}\mbox{{\it e-mail address:}} &
\mbox{Fadin@inp.nsk.su}\\
^{\dag\dag}\mbox{{\it e-mail address:}} &
\mbox{M.G.Kozlov@inp.nsk.su }\\
^{\ddag}\mbox{{\it e-mail address:}} &
\mbox{A.V.Reznichenko@inp.nsk.su}\\
\end{array}
$
\end{titlepage}

\section {Introduction}

Multi-Regge form of many-particle  amplitudes  underlies  the well-known  BFKL
(Balitsky--Fadin--Kuraev--Lipatov) approach \cite{Fadin:1975cb, Kuraev:1976ge,
Kuraev:1977fs, Balitsky:1978ic}, which gives the most common basis for the
description of small $x$  processes.  The idea of this form emerged in the process
of the calculations \cite{Lipatov:1976zz, Kuraev:1976ge}  of elastic scattering
amplitudes at large c.m.s. energies $\sqrt s$ and fixed momentum transfer
$\sqrt{-t}$ in the leading logarithmic approximation (LLA) which means  summation of
radiative corrections of the type of $(g^2\ln(s/|t|))^n$ ($g$ is the coupling
constant). The dispersive method  used in the calculations  requires knowledge of
all inelastic amplitudes in the multi-Regge kinematics (MRK) where produced
particles have limited (not growing with $s$) transverse momenta and strongly
ordered longitudinal momenta.  It turned out \cite{Lipatov:1976zz, Kuraev:1976ge}
that these amplitudes  have  the multi-Regge form in the first few orders of
perturbation theory. This led to the hypothesis that this form is valid in the LLA
in all orders of perturbation theory. Lately, this hypothesis has been  proved
\cite{Balitskii:1979}.  Then,  it  was generalized for the next-to-leading
logarithmic  approximation (NLLA),  which  means summation of radiative corrections
of the type of $g^2(g^2\ln(s/|t|))^n$. Note that in this approximation one has to
consider not only the LLA amplitudes  with $g^2$-corrections,   but also amplitudes
with a couple of particles having longitudinal momenta of the same order.  They
correspond to the kinematics which is called quasi multi-Regge (QMRK). To unify
consideration   we will use in the following the notion "jet"  both for such couple
of particles and  for a single particle and will treat  QMRK  as  MRK with jets.

The BFKL approach in the  NLLA  is widely used  in Quantum  Chromodynamics (QCD)
now. It is used  also in supersymmetric Yang-Mills theories (SYM);   in particular,
it was used in  the maximally extended (${\cal N}=4$) SYM for  check of
self-consistency of the ABDK-BDS (Anastasiou-Bern-Dixon-Kosower --
Bern-Dixon-Smirnov) ansatz \cite{Anastasiou:2003kj, Bern:2005iz} $M^{BDS}$ for
amplitudes with the maximal helicity violation  (MHV amplitudes) in the  multi-color
(planar) limit and for verification of the conjectures of dual conformal invariance
\cite{Drummond:2006rz, Bern:2006ew, Bern:2007ct, Alday:2007hr, Drummond:2007aua,
Drummond:2007cf, Nguyen:2007ya} and correspondence  between the MHV amplitudes and
expectation values of Wilson loops \cite{Drummond:2007aua,
Drummond:2007cf,Brandhuber:2007yx, Drummond:2007au,Drummond:2007bm,
Drummond:2008aq}, presentation of true amplitudes as  the product  $M^{BDS}$ on a
function of conformal-invariant ratios of kinematic invariants  $R$ called the
remainder factor, and  for the calculation of this factor in the multi-Regge
kinematics \cite{Bartels:2008ce, Bartels:2008sc,Lipatov:2010qg,
Lipatov:2010ad,Bartels:2010tx,Fadin:2011we,Fadin:2013hpa,Fadin:2014gra}.

To be confident in the results of the BFKL approach in the NLLA  one needs a proof
of validity of the multi-Regge form of many-particle amplitudes in this
approximation. The way of proving  based on $s$-channel unitarity was outlined  in
\cite{Fadin:2003av} and worked out in detail in \cite{Fadin:2006bj}.  The main steps
of the   proof are the following. The requirement of compatibility of the
$s$-channel unitarity with the Reggeized form of amplitudes leads to an infinite set
of  the relations (bootstrap relations) connecting derivatives of this form  over
energy variables  with the discontinuities in this variables, which, in turn, are
determined by this form. It turns out that all these relations are fulfilled  if
several conditions on the Reggeon  vertices and trajectory  (bootstrap conditions)
are valid. Thus, the proof of the multi-Regge form is reduced to check validity of
the bootstrap conditions.

In this paper we present the results necessary for this  check in  Yang-Mills
theories containing  fermions (we will call them also quarks) and scalars in
arbitrary representations   of the colour group  with a general form of the
Yukawa-type interaction. First, we define the multi-Regge form of  multiple
production amplitudes and present all components of this  form in the NLLA.  Then
the bootstrap approach to the proof of the validity  of this form  is sketched, all
main components of the bootstrap conditions are defined and  fulfilment of these
conditions  is discussed.

The paper  is organized as follows.  In  Section 2  we define  the multi-Regge form
of the MRK amplitudes and specify the theories in which this form will be proved. In
Section 3 we present the  Regge trajectory of the gauge boson (we call it gluon as
in QCD) and the  Reggeon  vertices entering in the  multi-Regge form.  In Section 4
the bootstrap approach is briefly presented and the  bootstrap conditions are
formulated. In Section 5 verification of these conditions is presented.

\section {The multi-Regge form of  multiple production  amplitudes}

\begin{figure}[t]
\centering
 \begin{minipage}{0.9\textwidth}%
 \epsfig{file=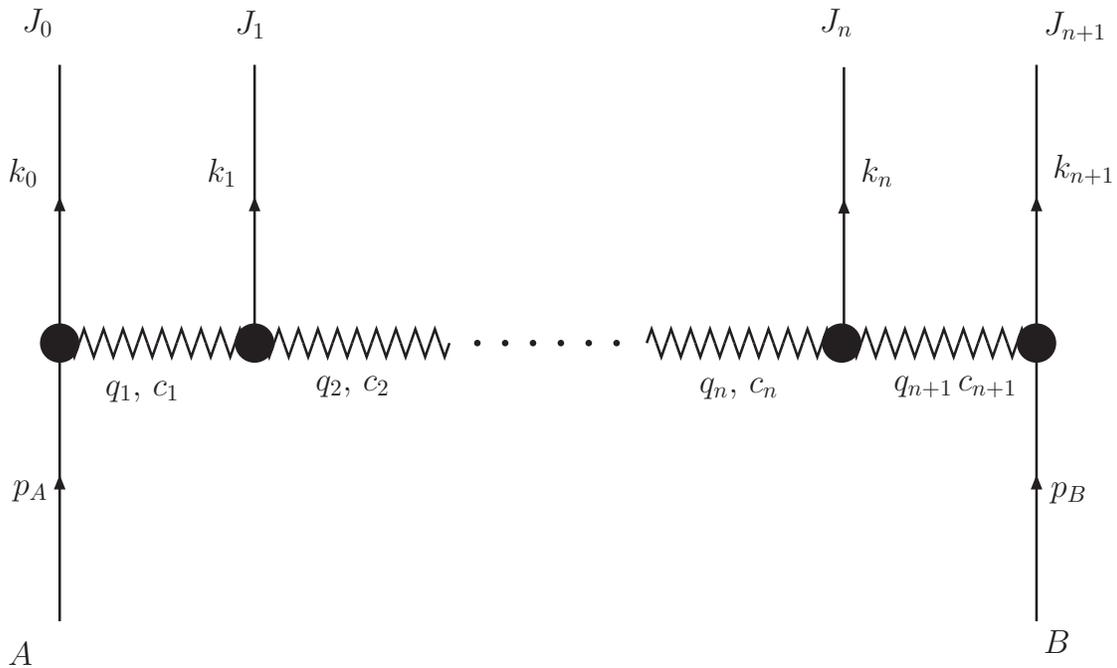,width=\textwidth}%
 \end{minipage} \\
 \caption{Schematic representation of the amplitude $2 \rightarrow 2+n$.  The zig-zag
 lines represent Reggeized gluon exchange. Right and left black blobs represent
 the Particle-Particle-Reggeon (PPR) vertices; black blobs in the middle represent the
 Reggeon-Reggeon-Particle  vertices.}
 \label{figmrk}
 \end{figure}

The multi-Regge form of the amplitude ${\cal A}_{2\rightarrow n+2}$ of   the process
$A+B\rightarrow J_0+J_1+\ldots+J_n+J_{n+1}$ is shown in Fig.\ref{figmrk}, where the
zig-zag lines represent Reggeized gluon (Reggeon) exchange,  right and left black
blobs represent  the Particle-Particle-Reggeon (PPR) vertices and  black blobs in
the middle represent the Reggeon-Reggeon-Particle (RRP) vertices. The PPR and RRP
vertices are called also scattering and production vertices correspondingly.

It is necessary to note here that the simple factorized form shown in
Fig.\ref{figmrk} is valid for the real parts of the MRK  amplitudes only. In fact,
the imaginary parts are much more complicated  than the real ones  and have not any
factorized form at all.

In the following for any   4-vector  $v$ we   use the decomposition
$v=v^{+}n_1+v^{-}n_2+v_{\perp}$ with light-cone vectors $n_{1,2}$ such that
$(n_1n_2)=1$, and therefore $v^{+} \equiv (v,n_2) ,\;\; v^{-}\equiv (v,n_1)$. It is
supposed that the dominant   components   of the  momenta $p_A$ and $p_B$ of the
initial particles $A$ and $B$ are $p^+_A$ and $p^-_B$ correspondingly, so that the
squared  energy  in the c.m.s. $s\simeq 2p^+_Ap^-_B$.  Each of the final jets
$J_i,\;\;  i=1, \ldots {n+1}$ with momentum $k_i=q_{i}-q_{i+1}\, \;\; q_0\equiv
p_A,\;\; q_{n+2}\equiv -p_B$ can represent either a single particle or a couple of
particles. Their  rapidities $y_i$, $y_i=\frac{1}{2}\ln\left(k^+_i/k^-_i\right)$ for
$i=1,\ldots n$,  $y_0= \ln\left(\sqrt 2 p^+_A/|q_{1\bot}|\right)$ and  $y_{n+1}=
\ln\left(|q_{(n+1)\bot}|/\sqrt 2 p^-_{B}\right)$  are  strongly ordered: $y_0 \gg
y_1 \gg \dots  \gg y_n \gg y_{n+1}$;  all $k_{i\perp}$ are limited.

In these denotations the multi-Regge form for the  real parts of the MRK amplitudes
can be  written as
\begin{equation}%
\Re {\cal A}_{2\rightarrow n+2}=2s\Gamma^{{R}_1}_{
J_0 A} \left( \prod_{i=1}^n
 \frac{e^{\omega(q_i)(y_{i-1}-y_i)}}{q^2_{i\perp}}\gamma^{J_i}_{{R}_i
 {R}_{i+1}}
 \right) \frac{e^{\omega(q_{n+1})(y_{n}-y_{n+1})}}{q^2_{(n+1)\perp}}
 \Gamma^{{R}_{n+1}}
_{J_{n+1} B},
\label{A 2-2+n}
\end{equation}
where $\omega(q)$ is called the  gluon trajectory (in fact, the trajectory is $1+\omega(q)$), ${\Gamma}^{{R}}_{ J_{0}A}$ and $\Gamma^{{R}} _{  J_{n+1} B}$ are the scattering vertices and  $\gamma ^{J_{i}}_{{R}_i {R}_{i+1}}$ are the production vertices.
The numerator of the Reggeon propagator ${e^{\omega(q_i)(y_{i-1}-y_i)}}=
\Big(s_i/\sqrt{k^2_{i-1\perp}k^2_{i\perp}}\Big)^{\omega(q_i)}$,
where $s_i= (k_i+k_{i+1})^2\approx 2 k^-_{i+1} k^+_{i}$   is known  as the Regge-factor.

In the NLLA  one has to  know the gluon  trajectory  with the  two-loop  accuracy,
the Reggeon vertices with  one-particle jets with the  one-loop corrections  and the
Reggeon vertices with  two-particle jets  at the Born approximation only.  In QCD
all these vertices and the trajectory were calculated  with the required accuracy
many years ago (see, for instance, \cite{Fadin:2003av} and references therein). Here
we present them for a wide class of Yang-Mills  theories with $n_f$ quark fields
$\psi_i^{a_i}$ ($a_i$ and $i$ are correspondingly  colour and flavour indices, $
i=1,\dots,\,n_f$) and $n_s$ (pseudo)scalar fields $\phi_r^{A_r}$ ($A_r$ and $r$ are
colour and flavour indices respectively, $ r=1,\dots,\,n_s$) in any
representations   of the colour group  with a general form of the Yukawa-type
interaction
\begin{equation}\label{lagrangian}
{\cal L}_Y=g_Y\left(\bar{\psi}_i^{a_i}[\gamma_5]_r\psi_j^{c_j}\right)\;
(R_{ij}^r)_{a_ic_j}^{b_r}\;\phi_r^{b_r}+h.c.\,.
\end{equation}
In the lagrangian (\ref{lagrangian}) $[\gamma_5]_r=1$ for scalars and
$[\gamma_5]_{r}=i\gamma_5=-\gamma^0\gamma^1\gamma^2\gamma^3$ for pseudoscalars;
$(R_{ij}^r)_{a_ic_j}^{b_r}$ are  flavour matrices of the Yukawa-type  interaction.
Different fields  transform according to different representations of the gauge
group $SU(N_c)$ with generators $T_{bc}^a=-if^{abc}$  for gluons,  $t_i^{a}$ for
quarks and ${\cal T}_i^{a}$ for scalars. The colour projectors
$(R_{ij}^r)_{a_ic_j}^{b_r}$ obey the  commutation relations following from the gauge
invariance:
\begin{equation}\label{eq:alg-R}
(t_f^{a})_{c_fb}(R_{fi}^r)_{bc_i}^{n_r}-(R_{fi}^r)_{c_fd}^{n_r}(t_i^{a})_{dc_i}=
(R_{fi}^r)_{c_fc_i}^{m_r}({\cal T}^a_r)_{m_r n_r}\,.
\end{equation}
Here the summation is only performed over colour indices  $b$, $d$, and $m_r$.
 We will use the symmetry factors $\kappa^f_i$ ($\kappa^s_r$) equal to ${1}/{2}$
for Majorana quarks (for the real scalars) and equal to 1 for Dirac quarks (for
complex scalars) and  the denotations
\begin{equation}
\xi_f=\sum_{i=1}^{n_f}\kappa^f_i\frac{T^f_i}{N_c}\,,
\;\xi_s=\sum_{r=1}^{n_s}\kappa^s_r\frac{T^s_r}{N_c}\,,
\end{equation}
where generators $T^f_i, \; T^s_r$  are normalized by the relations
\begin{equation}
\Tr[T^aT^b]=N_c\delta_{ab}\,, \Tr[t_i^at_i^b]=T^f_i\delta_{ab}\,,  \;\Tr[{\cal
T}_r^a{\cal T}_r^b]=T^s_r\delta_{ab}\,.
\end{equation}
Quadratic Casimir operators are defined as
\begin{equation}
T^aT^a = C_V=N_c, \;\;  t^a_it^a_i=C_F^i\,,\;{\cal T}_r^a{\cal T}_r^a=C_S^r.
\end{equation}
In the fundamental representation $T^f_i=1/2$ and $C_F^i=(N_c^2-1)/(2N_c)$. But
note that  we use  the denotations $t^a_i, T^f_i$ and $C_F^i$ for
any representation of the colour group for quarks. The quark loop
contributions  with the colour structure $\Tr[t^a_it^b_i]$ can be obtained from the
QCD ones (where quarks are in the fundamental representation) by the substitution
$n_f\rightarrow 2\sum_i\kappa^f_i T^f_i$, and the contributions with the colour
structure $t^a_it^b_it^a_i$ by the substitution $1/N_c^2\rightarrow 1-2C_F^i/N_c $ .
One can also restore  the contributions of  vacuum polarization by  scalars from
corresponding quark  contribution in QCD   by the substitution \cite{Fadin:2007xy,
Gerasimov:2010zzb} $n_f\rightarrow 2\sum_r\kappa^s_r T^s_r / (4(1+\epsilon))$.

It is worth noting that the interaction (\ref{lagrangian})  permits transitions  with nonconservation of fermion and scalar flavours. For the diagonal transitions we  omit the flavour  indices.

The $N$-extended SYM contains $n_M =N$ Majorana  quarks and $n_s =2(N-1)$ neutral
scalars. The matrices  $(R_{if}^r)_{a_ib_f}^{b_r}$ in  SYM have the form
$(R_{if}^r)_{a_ib_f}^{b_r}=\Delta_{if}^rT_{a_ib_f}^{b_r}$ and the  flavour matrices
$\Delta^r$ subject to the conditions  $[\Delta^r]^2=-1$, $\Tr[\Delta^r]=0$,
$\Tr[\Delta^r\Delta^t]=n_f\delta^{rt}$ with $n_f=n_M$. The Yukawa constant in SYM
reduces to $g_{Y}=g/2$.

As it is known,  dimensional regularization violates supersymmetry,
therefore  in SYM a modification of the dimensional regularization
is used  which is called dimensional reduction ~\cite{Siegel:1979wq}.
Hereafter to present explicitly    $N=4$  SYM results  we use the
dimensional reduction scheme, where $n_s=6-2\epsilon$.

\section {Gluon Regge trajectory and Reggeon vertices}

Apart from contributions of the Yukawa-type  interaction \eqref{lagrangian}, all
Reggeon vertices  as well as the gluon trajectory in the Yang-Mills  theories with
quarks and scalars in any representations  of the colour group    can be obtained
from known results  with the   NLLA accuracy by the substitutions  discussed above.

There are two kinds of scattering vertices:  with dominant $+$ and dominant $-$
components of particle momenta, or, in other words, in fragmentation region of
particles $A$ and $B$. Evidently, ones can be obtained from other by appropriate
substitutions.  We  present the scattering vertices for the particle $A$
fragmentation region.
\subsection{Gluon trajectory}
\label{subsection:{Gluon trajectory}} The two-loop  calculations of the trajectory
were carried out in Refs.
\cite{Fadin:1995dd,Fadin:1995xg,Fadin:1995km,Kotsky:1996xm,Fadin:1996tb} and then
confirmed in  \cite{Blumlein:1998ib,DelDuca:2001gu}. Using the integral
representation for the trajectory in  QCD \cite {Fadin:1995xg} we obtain in
$D=4+2\epsilon$ space time dimensions
\begin{equation}
\omega(-\vec{q}_{i}^{\;2})=\frac{-\bar{g}^{2}\;\vec{q}_{i}^{\;2}}
{\pi^{1+\epsilon}\Gamma(1-\epsilon)}\int\frac{d^{2+2\epsilon}k\;}
{\vec{k}^{\;2}(\vec{k}-\vec{q}_{i})^{2}}
\Biggl(1+\bar{g}^{2}\Biggl[f(\vec
{k},0)+f(0,\vec{k}-\vec{q}_{i})-f(\vec{k},\vec{k}-\vec{q}_{i})\Biggr]\Biggr)~,
\label{omega as double integral}
\end{equation}
where
\begin{equation}
\bar{g}^{2}
=\frac{g^{2}N_{c}\Gamma(1-\epsilon)}{(4\pi)^{2+{\epsilon}}}~,
\end{equation}
$\Gamma(x)$ is the Euler gamma-function, $g$ is the bare coupling,
and
\[
f(\vec{k}_{1},\vec{k}_{2})=\frac{(\vec{k}_{1}-\vec{k}_{2})^{2}}{\pi^{1+\epsilon}
\Gamma(1-\epsilon)}\int\frac{d^{2+2\epsilon}l}{(\vec{k}
_{1}-\vec{l})^{2}(\vec{k}_{2}-\vec{l})^{2}}\Biggl(\ln\left(  \frac{(\vec{k}_{1}-\vec{k}_{2})^{2}}{\vec{l}^{\;2}}\right)  -2\psi(1+2\epsilon)-\psi\left(1-\epsilon\right)
\]
\begin{equation}
+2\psi\left(1+\epsilon\right)  +\psi(1) -\frac{1}{\epsilon}
-\frac{a_1}{2(1+2\epsilon)(3+2\epsilon)}\Biggr)~,
\label{a omega}
\end{equation}
\be
\psi(x)=\frac{\Gamma^{\prime}(x)}{\Gamma(x)}, \;\; a_1=11+7\epsilon -4(1+\epsilon)\xi_f -\xi_s.
\ee
For $N=4$ SYM,  the coefficients $a_1$  vanishes in the dimensional reduction.

An explicit expression  for the trajectory was calculated   in QCD
\cite{Fadin:1996tb} only  the limit  $\epsilon\rightarrow 0$. Using this result we
obtain \be \omega(-\vec{q}^{\;2}) = -\bar g^2(\qs)^\epsilon
\frac{\Gamma^2(\epsilon)}{\Gamma(2\epsilon)} +\bar g^4(\qs)^{2\epsilon}\left[a_1
\left(\frac{1}{3\epsilon^2}-\frac{8}{9\epsilon} +\frac{52}{27} \right)
+\frac{2}{\epsilon}\zeta(2) -2\zeta(3) +{\cal O}(\epsilon)\right],  \label{omega
bare} \ee where $\zeta(n)$ is the Riemann zeta-function. In $N=4$ SYM with the
dimensional reduction  one has
\begin{equation}
\omega(-\vec{q}^{\;2})_{N=4
 SYM} =  -\bar g^2(\qs)^\epsilon \frac{\Gamma^2(\epsilon)}{\Gamma(2\epsilon)} +\bar g^4(\qs)^{2\epsilon}\left[\frac{2}{\epsilon}\zeta(2) -2\zeta(3) +{\cal O}(\epsilon)\right]
 ~. \label{omega SYM}
 \end{equation}
In the ${\overline{\mbox{MS}}}$ scheme the bare coupling is connected to the
renormalized coupling, $g_{\mu}$, through the relation
\begin{equation}
g=g_{\mu}\mu^{-\mbox{\normalsize $\epsilon$}}\left[  1+\bar{g}_{\mu}^{2}\frac
{\beta_0}{2N_c\epsilon}\right]  ~,\;\;\bar{g}_\mu^{2}
=\frac{g_\mu^{2}N_{c}\Gamma(1-\epsilon)}{(4\pi)^{2+{\epsilon}}}~, \;\;\frac
{\beta_0}{N_c}=\frac{11}{3}-\frac{4}{3}\xi_f-\frac{1}{3}\xi_s~.
\label{coupling renormalization}
\end{equation}

In terms of  the renormalized  coupling, one obtains
\[
\omega(-\vec{q}^{\;2}) = -\bar g^2_\mu \left(\frac{\qs}{\mu^2}\right)^\epsilon
\frac{\Gamma^2(\epsilon)}{\Gamma(2\epsilon)} -\bar
g^4_\mu\left(\frac{\qs}{\mu^2}\right)^{2\epsilon} \left[\frac{\beta_0}{N_c}
\left(\frac{1}{\epsilon^2}-\ln^2\left( \frac{\vec{q}^{\;2}}{\mu^2}\right)\right)+
\left(\frac{1}{\epsilon}+2\ln\left(\frac{\vec{q}^{\;2}}{\mu^2}\right)\right)\right.
\]
\begin{equation}
\left. \times \left(\frac{67}9-2\zeta(2)-\frac{20}9\xi_f-\frac{8}9\xi_s\right)
-\frac{404}{27}+2\zeta(3)+\frac{112}{27}\xi_f+\frac{52}{27}\xi_s+
{\cal O}(\epsilon)\right].  \label{omega expanded}
\end{equation}

\subsection{Vertices for one-particle jets}
Reggeon vertices with  gluons  are gauge invariant.  To simplify representation of
these vertices, we will use  for the polarization vector $e$ of the  gluon with the
momentum $k$ the light-cone gauge $(en_2)=0$, so that
\begin{equation}
e^\mu=e^\mu_\perp -\frac{(e, k)_{\perp}}{kn_2}n_2^\mu~. \label{n 2 gauge}
\end{equation}
It worth  noting that knowing some  vertex in this gauge, one  can restore its gauge
invariant  form. Here we have used the notation
$(a,b)_{\bot}\equiv(a_{\bot},b_{\bot})$.
\subsubsection*{Scattering vertices}
\label{Scattering vertices} Using results of Refs. \cite{Fadin:1993wh,
Fadin:1995xg,Fadin:1992zt, Kozlov:2014gaa} for the  one-loop gluon, quark and scalar
corrections correspondingly,  we obtain  for the {\bf gluon-gluon-Reggeon vertex
$\Gamma_{G'G}^{R}$}
\[
\Gamma_{G'G}^R=-g\bigl(e^{'*}, e\bigr)_{\bot}
T_{G'G}^R\Biggl[1-\bar{g}^2(-q_{\bot}^2)^{\epsilon}
\frac{\Gamma^2(1+\epsilon)}{\epsilon\Gamma(1+2\epsilon)}
\biggl(\frac{2}{\epsilon}+\psi(1)+\psi(1-\epsilon)-2\psi(1+\epsilon)-
\]
\[
-\frac{(1+\epsilon)^2 a_1+2\epsilon^2 a_2}{2(1+\epsilon)^2(1+2\epsilon)(3+
2\epsilon)}\biggr)\Biggr]-2g\bar{g}^2(-q_{\bot}^2)^{\epsilon}\frac{\Gamma^2(1+\epsilon)}
{(1+\epsilon)\Gamma(4+2\epsilon)}
\]
\begin{equation}
\times T_{G'G}^R e_{\bot\mu}^{'*}e_{\bot\nu}
\biggl(g_{\bot}^{\mu\nu}-(D-2)\frac{q_{\bot}^{\mu}q_{\bot}^{\nu}}{q_{\bot}^2}\biggr)a_2
, \label{eq:GRG-nlo}
\end{equation}
where $e$  and $e'$ are the polarization vectors  of  the gluons $G$   and $G'$
respectively, $q$ is the Reggeon   momentum, $T_{G'G}^R$ is the colour factor, \be
a_2=1+\epsilon -2\xi_f +\xi_s. \ee For $N=4$ SYM,  the coefficients $a_2$  vanishes
in the dimensional reduction.

The {\bf quark-quark-Reggeon vertex $\Gamma_{Q'Q}^R$ }  with one-loop accuracy was
calculated  in  QCD  in \cite{Fadin:1993qb}.  Scalar corrections in SYM were found
in \cite{Kozlov:2014gaa}.  Using these results, we obtain
\[
\Gamma_{Q'_fQ_i}^R=g\delta_{fi}\bar{u}_f(p')t_i^R\frac{\not{n}_2}{2p^+}u_i(p)\biggl[1-\bar{g}^2(-q_{\bot}^2)^{\epsilon}\frac{\Gamma^2(1+\epsilon)}{\epsilon\Gamma(1+2\epsilon)}\biggl(\frac{1}{\epsilon}+\psi(1-\epsilon)+\psi(1)-2\psi(1+\epsilon)
\]
\be +\frac{a_1-3(3+2\epsilon)}{2(1+2\epsilon)(3+2\epsilon)}
+\Bigl(\frac{2C_F^i}{N_c}-1\Bigr)\Bigl(\frac{1}{\epsilon}-\frac{3-2\epsilon}{2(1+2\epsilon)}\Bigr)\biggr)\biggr]+\Gamma_{Q'_fQ_i}^{R(Y)}\,,
\label{eq:QRQc}
\end{equation}
where $\Gamma_{Q'_fQ_i}^{R(Y)}$ is the contribution of  the Yukawa-type interaction.
We don't present it here in the general case  because  we don't need its explicit
form to  prove  the  validity of the bootstrap conditions.   In  SYM this term  is
absent due to the cancellation of the  scalar and pseudoscalar contributions. They
have  different signs because corresponding matrix elements contain odd numbers of
gamma matrices between two   matrices $\gamma_5$ in the  pseudoscalar case  and two
identity matrices in the scalar case.

The  {\bf scalar-scalar-Reggeon vertex $\Gamma_{S'_{r'}S_r}^R$}  was   calculated in
\cite{Kozlov:2014gaa} in SYM  by the method developed in \cite{Fadin:2001dc}. The
calculations can be easily extended to any  representation of the colour group with
the result
\begin{equation}
\begin{split} \label{eq:SRSc}
&\Gamma_{S'_{r'}S_r}^R=g \delta_{r'r}({\cal T}_r^R)_{S'_{r'}S_r}\Biggl[1-\bar{g}^2(-q_{\bot}^2)^{\epsilon}\frac{\Gamma^2(1+\epsilon)}{\epsilon\Gamma(1+2\epsilon)}\biggl(\frac{1}{\epsilon}+\psi(1-\epsilon)+\psi(1)-2\psi(1+\epsilon)+\\
&+\frac{a_1-4(3+2\epsilon)}{2(1+2\epsilon)(3+2\epsilon)}+\Bigl(\frac{2C_S^r}{N_c}-1\Bigr)\biggl[\frac{1}{\epsilon}-\frac{2}{1+2\epsilon}\biggr]\biggl)\Biggl]  +\Gamma_{S'_{r'}S_r}^{R(Y)}\,,
\end{split}
\end{equation}
where $\Gamma_{S'_{r'}S_r}^{R(Y)}$ is the contribution  of the Yukawa-type
interaction.  As well as for the quark vertex, we don't present it here in the
general case, because  we don't need its explicit form.  In  SYM we have
\cite{Kozlov:2014gaa} \be \Gamma_{S'S}^{R(Y)}=- g
T_{S'S}^{R}\bar{g}^2(-q_{\bot}^2)^{\epsilon}\frac{\Gamma^2(1+\epsilon)}{\epsilon\Gamma(1+2\epsilon)}\;
2\xi_f\frac{(-1)^{I_s}}{1+2\epsilon}\;, \ee where $I_s=0$  if  $S$ is a  scalar and
$I_s=1$ if  $S$ is a  pseudoscalar.

\subsubsection*{Production vertex}
\label{subsubsection:{Production vertex}} In the Born approximation the {\bf
Reggeon-Reggeon-gluon   vertex $\gamma_{R_1R_2}^G$} was obtained in
\cite{Lipatov:1976zz}. One-loop gluon corrections to the vertex were calculated  in
Refs. \cite{Fadin:1993wh,Fadin:1996yv,DelDuca:1998cx,Fadin:2000yp}. In the last
paper they were  obtained at arbitrary $D=4+2\epsilon$.  With the same accuracy, the
quark  and scalar corrections  were obtained in \cite{Fadin:1994fj} and
\cite{Gerasimov:2010zzb} respectively. At arbitrary $D$ the corrections are rather
complicated (mainly because of the gluon contribution). We present them here in the
form where only terms singular at small gluon transverse momentum $\vec k$
($k=q_1-q_2,\; q_{1,2}$ are the momenta of the Reggeons $R_{1,2}$)    are given  at
arbitrary D, but the other terms in the limit $\epsilon \rightarrow 0$.
\begin{equation}
\label{eq:RRG_nlo}
\begin{split}
&\gamma_{R_1R_2}^G=\gamma_{R_1R_2}^{G(B)}+2g\bar{g}^2T_{R_1R_2}^{G}e_{\bot\mu}^{*}(k)
q_{1\bot}^2V^{\mu}(q_1,q_2)~,
\end{split}
\end{equation}
where
\be
\gamma_{R_1R_2}^{G(B)}=-2gT_{R_1R_2}^Ge_{\bot\mu}^{*}
\biggl({q_{1\bot}^{\mu}}-k_{\bot}^{\mu}\frac{q_{1\bot}^2}{k_{\bot}^2}\biggr) \label{gamma B}
\ee
is the
Born vertex~\cite{Lipatov:1976zz} in the light-cone gauge $(e,n_2)=0$,
\[
V^{\mu}(q_1,q_2)=
 \Bigl(\frac{11}{6}-\frac{2\xi_f}{3}-\frac{\xi_s}{6}\Bigr)
\Biggl(\frac{k^{\mu}_{\bot}}{k_{\bot}^2}-\frac{q_{1\bot}^{\mu}}{q_{1\bot}^2}\frac{q_{1\bot}^2+q_{2\bot}^2}{q_{1\bot}^2-q_{2\bot}^2} \Biggr)\ln\frac{q_{1\bot}^2}{q_{2\bot}^2} +
\Bigl(\frac{1}{6}-\frac{\xi_f}{3}+\frac{\xi_s}{6}\Bigr)
\]
\[
\times\Biggl[ \Biggl(\biggl(\frac{k^{\mu}_{\bot}}{k_{\bot}^2}-\frac{q_{1\bot}^{\mu}}{q_{1\bot}^2}\biggr)
\frac{2k_{\bot}^2}{\bigl(q_{1\bot}^2-q_{2\bot}^2\bigr)^2}
+\frac{k_{\bot}^{\mu}
\bigl(2k_{\bot}^2-q_{1\bot}^2-q_{2\bot}^2\bigr)}{q_{1\bot}^2
\bigl(q_{1\bot}^2-q_{2\bot}^2\bigr)^2}\biggr)\Bigl[q_{1\bot}^2+q_{2\bot}^2-
\frac{2q_{1\bot}^2q_{2\bot}^2}{q_{1\bot}^2-q_{2\bot}^2}\ln\frac{q_{1\bot}^2}
{q_{2\bot}^2}\Bigr]- \frac{k_{\bot}^\mu}{q_{1\bot}^2}
\Biggr]
\]
\be
-\frac12\biggl(\frac{k^{\mu}_{\bot}}{k_{\bot}^2}-\frac{q_{1\bot}^{\mu}}{q_{1\bot}^2}\biggr)
\biggl(\ln^2\frac{q_{1\bot}^2}{q_{2\bot}^2}+\frac{2\bigl|k_{\bot}^2\bigr|^{\epsilon}}{\epsilon^2}-{\pi^2} \biggr)~.\label{V mu}
\ee
For $N=4$ SYM in the  dimensional reduction scheme
\begin{equation}
\gamma_{R_1R_2}^{G} =\gamma_{R_1R_2}^{G(B)}\biggl(1-\bar{g}^2\biggl[ \frac{[-k_{\bot}^2]^{\epsilon}}{\epsilon^2}-\frac{\pi^2}{2}+ \frac{1}{2}\ln^2\Big[\frac{q_{1\bot}^2}{q_{2\bot}^2}\Big]\biggr]\biggr).
\end{equation}

\subsection{Vertices for two-particle jets}
Now we turn to vertices which are absent in the LLA and  appear in the NLLA. They
are needed in the Born approximation only.

\subsubsection*{Scattering vertices}
We will present the vertices  $\Gamma_{JP}^R$ of the   transition of a particle $P$
to  a two-particle jet $J$. The vertex  of the inverse transition $\Gamma_{PJ}^R
=\left(\Gamma_{JP}^R\right)^*$. We denote the momentum of the initial particle $k$
and the momenta  of the  final particles $l_1,\,l_2$, total jet momentum is
$l=l_1+l_2,\; l^+=k^+$ (remind, we are in the particle $A$ fragmentation region),
\begin{equation}
k=k^+n_1-\frac{k_{\bot}^2}{2k^+}n_2+k_{\bot}\,,\quad l_i=x_il^+n_1-\frac{l_{i\bot}^2}{2x_il^+}n_2+l_{i\bot},\;i=1,\,2, \;x_1+x_2 =1.
\end{equation}

The vertex of  { \bf {quark $\rightarrow$  quark-gluon  jet} transition $\Gamma_{\{Q
G \}Q}^R$ }  has the same form as in QCD    \cite{Fadin:1995km,Fadin:1999df}. It can
also  be written as in \cite{Ioffe:2010zz}:
\begin{equation}\label{eq:q-qg}
\begin{split}
&\Gamma_{\{Q G \}Q}^R=g^2e^*_{\bot\mu}\bar{u}(l_1)\frac{\not{n}_2}{2k^+}
\Bigl[t^G_it^R_i\bigl({\cal A}^\mu_b(x_2l_{1\perp}-x_1l_{2\perp})-{\cal A}^\mu_b(l_{1\perp}-x_1k_{\perp})\bigr)
\\
&-t^R_it^G_i\bigl({\cal A}^\mu_b(-l_{2\perp}+x_2k_{\perp})-{\cal A}^\mu_b(l_{1\perp}-x_1k_{\perp})\bigr)
\Bigr]u(k)~,
\end{split}
\end{equation}
where $e$ is the gluon polarization vector,   quark  colour and flavour  wave
functions are included in  $\bar{u}(l_1)$ and $u(k)$,
\begin{equation}
{\cal A}^\mu_b(p)=-\frac{1}{p^2}\left(x_1\gamma^\mu {\not{p}}+{\not{p}\gamma^\mu }\right)~.
\label{A b mu}
\end{equation}

Let us present the vertices of the gluon $G$ transition to pairs $\{ P_1(l_1), \bar P_2(l_2)\}$ in the form
\begin{equation}\label{A P1 P2}
\Gamma_{\{ P_1 \bar P_2 \}G}^R =g^2 e_{\bot\mu}\bigl(\textbf{T}^{G}\textbf{T}^{R}
A^{\mu}_{P_1P_2}(k)+\textbf{T}^{R}\textbf{T}^{G}A^{\mu}_{P_2P_1}(k)\bigr)\,,
\end{equation}
where ${\mathbf T}^R$ are the colour group generators for produced particles
in the corresponding representation. Generators $\Big(\textbf{T}^{R}\textbf{T}^{G}\Big)_{P_1 P_2}$
and $\Big(\textbf{T}^{G}\textbf{T}^{R}\Big)_{P_1 P_2}$ operate with the colour wave functions
of the particle produced in (\ref{A P1 P2}).

For {\bf gluon $\rightarrow$ quark-antiquark transition} one has
\be A^\mu_{Q\bar
Q}(k) =\bar{u}(l_1)\frac{\not{n}_2}{2k^+}\bigl({\cal
A}^\mu_p(l_{1\perp}-x_1k_{\perp})-{\cal A}^\mu_p(x_2l_{1\perp}-x_1l_{2\perp})\bigr)
v(l_2)~, \label{eq:g-q bar q}
\ee with
\begin{equation}
{\cal A}^\mu_p(p)=\frac{1}{p^2}\left(x_1\gamma^\mu {\not{p}}-x_2{\not{p}\gamma^\mu
}\right)~. \label{A p mu}
\end{equation}
The  second term in (\ref{A P1 P2}) reads as follows (the minus sign is associated with Fermi statistics)
\be
A^\mu_{\bar Q Q}(k) = - A^\mu_{ Q  \bar Q}(k)\Bigl|_{l_1 \leftrightarrow l_2}=
-\bar{u}(l_1)\frac{\not{n}_2}{2k^+}
\bigl({\cal A}^\mu_p(-l_{2\perp}+ x_2k_{\perp})-{\cal
A}^\mu_p(x_2l_{1\perp}-x_1l_{2\perp})\bigr) v(l_2)~.
\ee

Let us note that the vertex $\Gamma_{\{Q \bar Q \}G}^R$   \cite{Fadin:1999de}  can be
obtained from $\Gamma_{\{Q G \}Q}^R$  by crossing, i.e. by the replacement \be
x_2\rightarrow \frac{1}{x_2}, \; x_1\rightarrow -\frac{x_1}{x_2},\;
l_{2\bot}\leftrightarrow - k_\bot, \; e^*_{\bot\mu}\rightarrow e_{\bot\mu},  \;
u(k)\rightarrow v(l_2)~. \label{crossing} \ee


The gauge invariant  {\bf gluon $\rightarrow$  gluon-gluon jet  Reggeon vertex
$\Gamma_{\{G_1G_2  \}G}^R$ }  was obtained in  \cite{Fadin:1989kf}. In the
light-cone gauge we have \cite{Fadin:1999de} for the representation (\ref{A P1 P2}):

\be \label{eq:grgg} A^\mu_{G_1G_2}(k) = 2
e^{*\nu}_{1\bot}e^{*\rho}_{2\bot}\bigl({\cal A}_{\mu\nu\rho}(l_{1\bot} -
x_1k_{\bot})-{\cal A}_{\mu\nu\rho}(x_2l_{1\bot} - x_1l_{2\bot})\bigr)~,  \ee
where $e_{1,2}$ are the polarization vectors  of the gluons $G_{1,2}$ with the momenta
$l_{1,2}$, and
\begin{equation}
{\cal A}_{\mu\nu\rho}( p)=\frac{1}{p^2}\bigl(x_1x_2 g^{\nu\rho} p^\mu-x_1
g^{\mu\nu}p^\rho-x_2 g^{\mu\rho}p^\nu \bigr)~. \label{M mu nu rho}
\end{equation}


For the vertex  of the  {\bf scalar pair $\{S(l_1),\,\bar{S}(l_2)\}$ production by the
gluon $G(k)$} \cite{Kozlov:2014gaa} we have in (\ref{A P1 P2})
\be
\label{eq:grss}
A^\mu_{S\bar S}(k)=-2\Big(M_p^{\mu}(l_{1\bot}  -
x_1k_{\bot})-M_p^{\mu}(x_2l_{1\bot} - x_1l_{2\bot} )\Big)~,
\ee
where
\begin{equation}
M_p^{\mu}(p)=x_1 x_2 \frac{p^\mu}{p^2}~. \label{M p mu}
\end{equation}


The  {\bf scalar $\rightarrow$ scalar-gluon jet}   vertex   can be easily obtained from the previous one
by  the crossing replacement \eqref{crossing}:
\begin{align}\label{eq:SRSG}
&\Gamma_{\{GS'\}S}^R=-2g^2e_{\bot\mu}^*\biggl[\Big({\cal T}^G_r{\cal T}^R_r\Big)_{S'S}\Big( M_b^{\mu}(x_2l_1-x_1l_2)-M_b^{\mu}(l_1-x_1k)\Big)\\
&-\Big({\cal T}^R_r{\cal T}^G_r\Big)_{S'S}\Big(M_b^{\mu}(-l_2+x_2k)-M_b^{\mu}(l_1-x_1k)\Big)\biggr],\; M_b^{\mu}(p)=x_1\frac{p_{\bot}^{\mu}}{p_{\bot}^2}\,.\notag
\end{align}

The rest  {\bf particle $\rightarrow $ two-particle jet}  transitions exist due to
Yukawa-type interaction. The Reggeon  vertices for these transitions in SYM  were
calculated in \cite{Kozlov:2014gaa}.  The {\bf scalar $\rightarrow $ quark-antiquark}
vertex is written as
\begin{equation}\label{eq:ver:s-qq}
\begin{split}
\Gamma_{\{Q_i\bar{Q}_j\}S_r}^R&=-gg_Y\bar{u}_{i}(l_1)\frac{\not{n}_2}{2k^+}\Biggl[t^R_{i}(R_{ij}^r)^{S_r}\biggl( \frac{(x_2\not{l}_1-x_1\not{l}_2)_{\bot}}{(x_2l_1-x_1l_2)_{\bot}^2}+ \frac{(\not{l}_2-x_2\not{k})_{\bot}}{(l_2-x_2k)_{\bot}^2}\biggr)+\\
+&(R_{ij}^r)^{S_r}t_{j}^R\biggl(\frac{(x_1\not{l}_2-x_2\not{l}_1)_{\bot}} {(x_2l_1-x_1l_2)_{\bot}^2}+\frac{(\not{l}_1-x_1\not{k})_{\bot}} {(l_1-x_1k)_{\bot}^2}\biggr)\Biggr][\gamma_5]_rv_{j}(l_2)\,,
\end{split}
\end{equation}
where $i$ is quark and  $j$ is and anti-quark flavour, $r$ is the scalar flavour
($S_r$ is the scalar colour index); $[\gamma_5]_r=1$ if $S$ is the scalar, and
$[\gamma_5]_r=i\gamma_5$ for the pseudoscalar case. The crossing vertex
$\Gamma_{\{Q'(l_1)S(l_2)\}Q(k)}^R$ is
\begin{align}\label{eq:ver:q-qs}
\Gamma_{\{Q'_iS_r\}Q_j}^R&=-gg_Y\bar{u}_{i}(l_1)\frac{\not{n}_2}{2k^+}x_2 \biggl[t^R_i\Bigl[(R_{ji}^{r})^{S_r}\Bigr]^{\dag}\biggl(\frac{(\not{l}_1-x_1\not{k})_{\bot}}{(l_1-x_1k)_{\bot}^2}+ \frac{(\not{l_2}-x_2\not{k})_{\bot}}{(l_2-x_2k)_{\bot}^2}\biggr)+\\
+&\Bigl[(R_{ji}^{r})^{S_r}\Bigr]^{\dag}t^R_j\biggl(\frac{(x_2\not{l}_1-x_1\not{l}_2)_{\bot}}
{(x_2l_1-x_1l_2)_{\bot}^2}-\frac{(\not{l}_1-x_1\not{k})_{\bot}}
{(l_1-x_1k)_{\bot}^2}\biggr)\biggr][\gamma_5]_r u_{j}(k)\,.\notag
\end{align}

\subsubsection*{Production vertices}
\label{subsubsection:{Production vertices}}

Denoting momenta of produced particles $P_1$ and $P_2$ as $l_1$ and  $l_2$, of
Reggeons $R_1,\,R_2$ momenta as $q_{1}$  and $q_{2}$,  $q_1-q_2 = l_1+l_2=l$,  we
have for the jet production vertices:
\begin{equation}\label{eq:rrpp}
\gamma_{R_1R_2}^{\{P_1P_2\}}=g^2\bigl(\textbf{T}^{R_1}\textbf{T}^{R_2}B_{P_1P_2}(q_1;l_1,l_2)+\textbf{T}^{R_2}\textbf{T}^{R_1}B_{P_2P_1}(q_1;l_2,l_1)\bigr)\,,
\end{equation}
where ${\mathbf T}^R$ are the colour group generators for produced particles. Here
for quark-antiquark production $\{Q (l_1), \bar{Q} (l_2)\}$ one has
\cite{Fadin:1996nw,Fadin:1997hr,Fadin:1999yv}
\begin{equation}\label{eq:Bqq}
\begin{split}
&B_{Q \bar Q}(q_1; l_1, l_2) =\bar{u}(l_1)\frac{\not{n}_2}{l^+}
b(q_1;l_1,l_2)v(l_2)~,\\
&\;B_{\bar{Q} Q}(q_1; l_2, l_1)= - B_{Q \bar Q}(q_1; l_1, l_2) \Bigl|_{l_1 \leftrightarrow l_2}
=-\bar{u}(l_1)\frac{\not{n}_2}{l^+} \overline{b(q_1;l_2,l_1)}v(l_2)\,,
\end{split}
\end{equation}
where
\begin{equation}\label{eq:b-q}
\begin{split}
&b(q_1;l_1,l_2)=\frac{\notp[1]{l}(\notp[1]{l}-\notp[1]{q})}{x_1(q_1-l_1)_{\bot}^2
+x_2l_{1\bot}^2}+\frac{x_1x_2}{\Lambda^2_{\bot}}\biggl[\frac{q_{1\bot}^2\bigl(
\notp[1]{l}\notp{\Lambda}-\notp{\Lambda}\notp[2]{l}\bigr)}{\Lambda_{\bot}^2+x_1x_2
l_{\bot}^2}+\frac{\notp{\Lambda}\notp[1]{q}}{x_1}-\frac{\notp[1]{q}
\notp{\Lambda}}{x_2}\biggr]-1~,\\
&\overline{b(q_1;l_1,l_2)}=\gamma^0b^{\dag}(q_1;l_1,l_2)\gamma^0\,,\qquad\Lambda_{\bot}^{\mu}=\bigl(x_2l_{1}-x_1l_{2}\bigr)^{\mu}_{\bot}\,.
\end{split}
\end{equation}

For the  vertex of two-gluon  $\{G_1 (l_1), G_2 (l_2)\}$ production the result was
obtained in the  gauge invariant form \cite{Fadin:1989kf}. In the light-cone gauge
\eqref{n 2 gauge} it reads as \cite{Fadin:2003xs}:
\begin{equation}\label{eq:Bgg}
\begin{split}
&B_{G_1 G_2}(q_1; l_1,l_2)=4e^{*\alpha}_{1\bot}e^{*\beta}_{2\bot}\Biggl(\frac{1}{2}g^{\alpha\beta}_{\bot}\Biggl[
\frac{x_1x_2}{\Lambda_\bot^2}\Bigl(-2(q_{1\bot},\Lambda_\bot)
+q_{1\bot}^2\frac{\left(\Lambda_\bot, x_2l_{1\bot}+x_1l_{2\bot}\right)}
{x_2l^2_{1\bot}+x_1l^2
_{2\bot}}\Biggr)\\
&-x_1x_2\frac{q_{1\bot}^2-2(q_{1\bot},l_{1\bot})}{x_1(q_1-l_1)^2_{\bot}
+x_2l_{1\bot}^2}\Biggr]-\frac{x_2l_{1\bot}^{\alpha}q_{1\bot}^{\beta}
-x_1q_{1\bot}^{\alpha}(q_1-l_1)_{\bot}^{\beta}}{x_1(q_1-l_1)^2_{\bot}
+x_2l_{1\bot}^2}-\frac{x_1q_{1\bot}^{2}l_{1\bot}^{\alpha}(q_1-l_1)^{\beta}_{\bot}}
{l_{1\bot}^2(x_1(q_1-l_1)^2_\bot +x_2l_{1\bot}^2)}\\
&+\frac{x_1q_{1\bot}^{\alpha}\Lambda_{\bot}^{\beta}+x_2q_{1\bot}^{\beta}
\Lambda_{\bot}^{\alpha}}{\Lambda_\bot^2}+\frac{x_1q_{1\bot}^2l_{1\bot}^{\alpha}
l_{2\bot}^{\beta}}{l_{1\bot}^2(x_2l^2_{1\bot}+x_1l^2
_{2\bot})}-\frac{x_1x_2q_{1\bot}^2}{\Lambda_{\bot}^2(x_2l^2_{1\bot}+x_1l^2
_{2\bot})}\Bigl(\Lambda_{\bot}^{\alpha}
l_{2\bot}^{\beta}+l_{1\bot}^{\alpha}\Lambda_{\bot}^{\beta}\Bigr)\Biggr)\,.
\end{split}
\end{equation}


For the vertex of two scalar $\{S(l_1) \bar S  (l_2)\}$ production one has
\cite{Gerasimov:2010zzb}:
\begin{align}\label{eq:Bss}
&B_{S\bar S }(q_1; l_1, l_2)=2q^2_{1\perp}x_1x_2\Bigg\{\Bigg[\frac{x_2-x_1}{(l_1-x_1 q_1)^2_{\perp} +x_1x_2q^2_{1\perp}}+2\frac{(q_1,\Lambda )_{\perp}}{q^2_{1\perp}\Lambda^2_{\perp}}-\\
&-2\frac{(q_1,l_1-x_1q_1)_{\perp}}{q^2_{1\perp}[(l_1-x_1 q_1)^2_{\perp}
+x_1x_2q^2_{1\perp}]}\Bigg]-\Bigg[q_1 \rightarrow l\Bigg]\Bigg\}.\notag
\end{align}

\section {Bootstrap approach to the proof of the multi-Regge amplitude form}

\subsection{Bootstrap relations}

In QCD, the    scheme of the  proof  was formulated in  \cite{Fadin:2006bj}. The
main point of the scheme is use of the restrictions imposed on the amplitudes with
negative signatures in all $t_i$-channels  by   the unitarity conditions.

Signature (positive or negative) is a quantum number  attributed to  Reggeons  in
the  theory of complex angular momenta. Amplitudes with Reggeon exchanges have
corresponding signatures. At high energy  it means the corresponding  symmetry with
respect to the sign change of the energy variables.   The signature of the Reggeized
gluon is negative, i.e. the MRK amplitudes with the Reggeized gluon exchange in the
channel $t_i$ are odd with  respect to the replacements $s_{jk}\rightarrow -s_{jk},
\; (s_{jk})=(k_i+k_j)^2$ for $k\ge i \ge j+1$.  For the MRK amplitudes in the Born
approximation this property is fulfilled for any $t_i$ thanks to the  common factor
$s$. Therefore the  Born amplitudes have negative signatures in all $t_i$--channels
and can be considered as amplitudes with Reggeized gluon exchanges in all these
channels. In higher approximations conventional amplitudes are  given by a sum of
amplitudes with definite signatures (they are called signaturized amplitudes) in
some set of the $t_i$--channels over all sets and over positive and negative
signatures in each channel.  But the leading contribution  is given by the
amplitudes with negative signatures in all the $t_i$--channels. Indeed, due to the
negative signature of the Born amplitudes the symmetry  of the radiative corrections
is opposite to the signature of the amplitudes. It leads to cancellation of the
leading logarithmic terms in the amplitudes with the positive signatures. The
amplitudes with the positive signature even in one of the $t_i$-channels loose at
least one power of logarithm in the imaginary part and two powers in the real part.
Therefore with the NLLA accuracy the real part of the conventional amplitude
presented in \eqref{A 2-2+n} coincides with the real part of the amplitude ${\cal
A}^{\{-\}}_{2\rightarrow 2+n}$ with the Reggeized gluons (i.e. with the negative
signatures) in all the $t_i$ channels, $\Re {\cal A}_{2\rightarrow 2+n} =\Re{\cal
A}^{\{-\}}_{2\rightarrow 2+n}$.

According to the Steinmann theorem \cite{S:1960} on  absence of simultaneous
singularities of amplitudes in overlapping channels (two channels $s_{i_1,j_1}$ and
$s_{i_2,j_2}$ are called  overlapping if either $i_1<i_2\leq j_1< j_2$ or
$i_2<i_1\leq j_2< j_1$), the amplitude ${\cal A}^{\{-\}}_{2\rightarrow 2+n}$ can be
presented as a sum of contributions corresponding to various sets of the $n+1$
non-overlapping channels  \cite{Bartels_1,Bartels_2}.  Each of the contributions is
a  series in logarithms of independent energy variables $s_{i_k,j_k}$ of the
non-overlapping channels   symmetrized with respect to simultaneous change of signs
of all $s_{i,j}$ with $i < k \le j$, performed independently for each $k = 1, . . .
, n+1$, with the  coefficients which are  real functions of transverse momenta.
Using the equality \be \frac{\mathrm{disc}_{s}[\ln^r(-s)+\ln^r(s)]}{-\pi i}
=\frac{\partial}{\partial\ln s}\Re[\ln^r(-s)+\ln^r(s)] \label{identity} \ee valid
with the NLLA accuracy, one can obtain, with the same accuracy,  the "differential
dispersion relation" \cite{Fadin:2002et}:
\begin{equation}
\frac{1}{-\pi i}\left(\sum_{l=j+1}^{n+1}\mathrm{disc}_{s_{j,l}}
-\sum_{l=0}^{j-1}\mathrm{disc}_{s_{l,j}}\right){\cal A}^{\{-\}}_{2\rightarrow 2+n}\Big/s \,=\,
\frac{\partial }{\partial y_{j}}\; \left(\Re\,{\cal A}^{\{-\}}_{2\rightarrow 2+n}\Big/s\right)\,.
\label{deriv}
\end{equation}
which permit to  express the partial derivatives  ${\partial }/{\partial y_{j}}$ of
the real parts of the amplitudes ${\cal A}^{\{-\}}_{2\rightarrow 2+n}$ (divided by
$s$)  in terms of their discontinuities.  The important point here is that with the
NLLA accuracy the discontinuities themselves can be calculated using the real parts
of  ${\cal A}_{2\rightarrow 2+n}$ in the unitarity conditions in the $s_{i,j}$
channels.  On the other hand, the derivatives ${\partial }/{\partial y_{j}}$
determine  dependence of ${\Re\,\cal A}_{2\rightarrow 2+n}/s$ on $\ln s_{i,j}$, so
that using \eqref{deriv} one can  restore ${\Re\cal A}_{2\rightarrow 2+n}/s$
unambiguously  order by order in powers of $\ln s_{i,j}$  starting from the initial
conditions (in the NLLA  these conditions include, besides  the  tree   amplitudes,
one loop  amplitudes at some energy scale).

As it was explained  before, with the NLLA accuracy   ${\Re\,\cal
A}^{\{-\}}_{2\rightarrow 2+n}$  can be replaced by $\Re \,{\cal A}_{2\rightarrow
2+n}$, where ${\cal A}_{2\rightarrow 2+n}$ is the conventional amplitude. Assuming
that  ${\Re \,\cal A}_{2\rightarrow 2+n}$  \eqref{deriv} in the right part of
\eqref{deriv} has the multi-Regge form \eqref{A 2-2+n}, we come to the relations
(which are called \emph{bootstrap relations})
\begin{equation}
\frac{1}{-\pi
i}\left(\sum_{l=j+1}^{n+1}\mathrm{disc}_{s_{j,l}}
-\sum_{l=0}^{j-1}\mathrm{disc}_{s_{l,j}}\right){\cal
A}^{\{-\}}_{2\rightarrow n+2} \, =\,
 \left(\omega(t_{j+1})-\omega(t_{j})\right)
{\Re\, \cal A}_{2\rightarrow n+2}~.\label{bootstrap relations}
\end{equation}
It follows from the foregoing  that fulfilment of these  relations with the
discontinuities in the left side calculated  using $\Re\, {\cal A}_{2\rightarrow
n+2}$ in the unitarity conditions ensures the Reggeized form of energy dependent
radiative corrections. Therefore, in order  to prove the validity of the multi-Regge
form \eqref{A 2-2+n} in the NLLA  (assuming that this form  is correct at some scale
in the one loop approximation)  enough  to prove  that  the bootstrap  relations
\eqref{bootstrap relations} are fulfilled.  At first glance, this problem seems
insoluble because of  the infinite number of these relations.  However,  it turns
out \cite{Fadin:2006bj}  that the infinite set of the bootstrap relations
\eqref{bootstrap relations}  is fulfilled if several nonlinear conditions (which are
called bootstrap conditions) imposed on the Reggeon vertices and the gluon
trajectory  hold true. This statement plays a crucial role   in the proof of the
correctness of the form \eqref{A 2-2+n}. It was proved in QCD using the operator
form of the discontinuities  \cite{Fadin:2006bj} in the left side of
\eqref{bootstrap relations}. The proof remains valid for  Yang-Mills  theories
containing  fermions and scalars in arbitrary representations of the colour group
with any Yukawa-type interaction despite of change of the fermion contributions and
appearance additional scalar contributions to the discontinuities.

\subsubsection*{Representation of the discontinuities}

The operator form is defined in  the space of states  $|{\cal G}_1{\cal G}_2\rangle$
of two $t$-channel Reggeons with the  orthonormality property \be \langle{\cal
G}'_1{\cal G}'_2|{\cal G}_1{\cal G}_2\rangle =\xs_1\xs_2
\delta(\x_1-\xp_1)\delta(\x_2-\xp_2)\delta_{{\cal G}_1{\cal G}'_1}\delta_{{\cal
G}_2{\cal G}'_2}~, \label{orthonormality} \ee where $\x_i$ and $\xp_i$ are the
Reggeon transverse momenta and ${\cal G}_i$ and ${\cal G}'_i$ are their colour
indices.  The main elements of this form  are the impact factors for particle-jet
and Reggeon-jet transitions and the operators of the BFKL kernel $\hat{\cal
K}(y_k-y_{k+1})$ and  the jet production. Remind that we use the notion jet  both
for a single particle and  for a  couple of particles  having   longitudinal momenta
of the same order.  As an example,  let us present  the discontinuity of ${\cal
A}^{\{-\}}_{2\rightarrow n+2}$  in $s_{k,n+1}$-channel  (for a schematic
representation of this discontinuity see Fig. \ref{figexample}):

 \begin{figure}[t]
\begin{center}
 \begin{minipage}{0.75\textwidth}%
 \epsfig{file=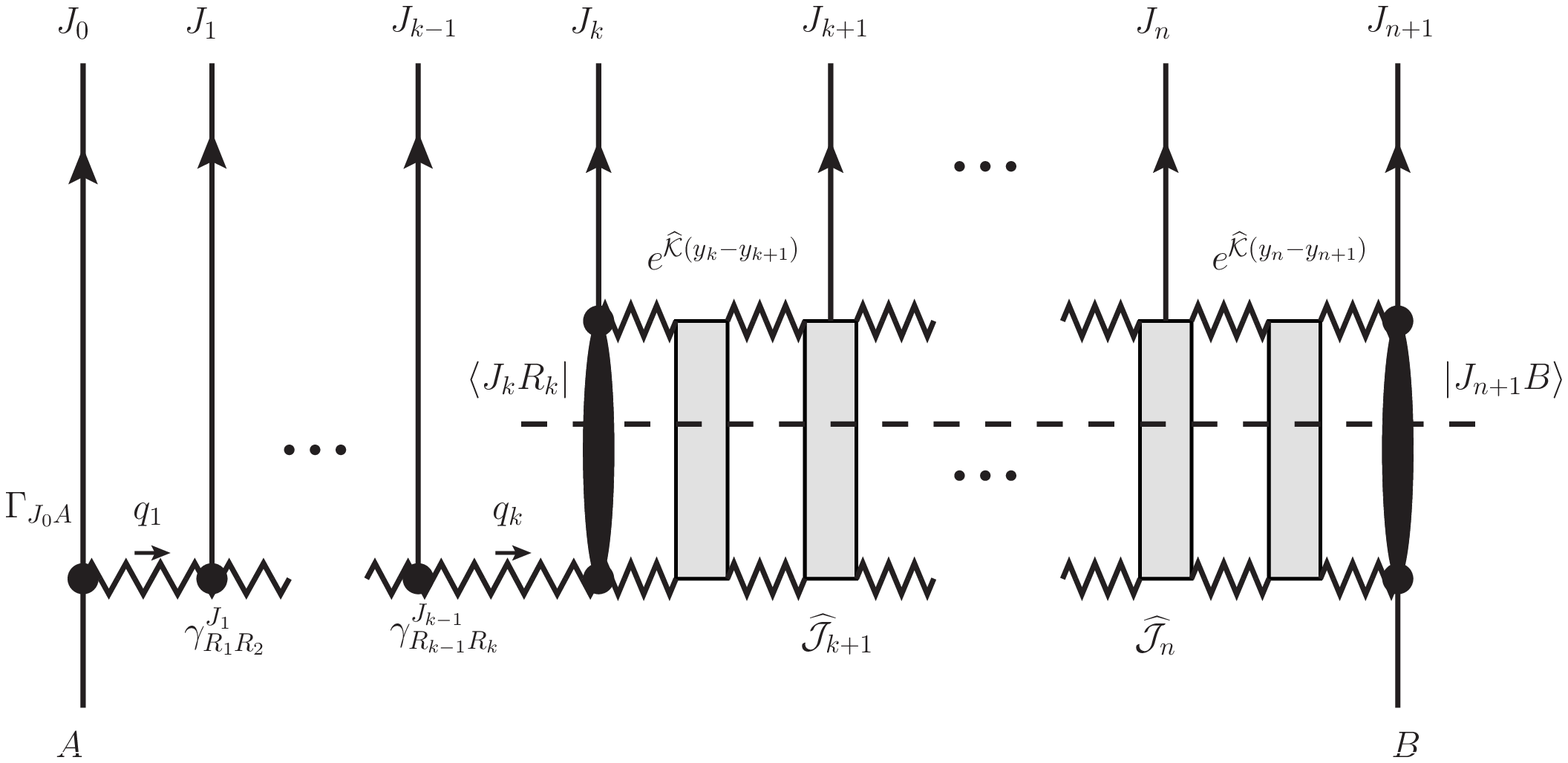,width=\textwidth}%
 \end{minipage} \\
 \end{center}
 \caption{Schematic representation of the discontinuity of ${\cal
 A}^{\{-\}}_{2\rightarrow n+2} $  in $s_{k,n+1}$-channel}
The zig-zag lines represent  Reggeized gluon exchange.    The dashed line denotes
on-mass-shell states in the unitarity condition. The right and the left black ovals
represent the impact-factors for particle-jet and Reggeon-jet transitions
respectively. The  right-angled grey blocks denote  the operators of the jet
production. The blank right-angled block in the  $t_{k+1}$-channel represents  the
operator  $e^{\hat{\cal K}(y_k-y_{k+1})}$.
 \label{figexample}
 \end{figure}

\begin{equation*}
\begin{split}
&-4i(2\pi)^{D-2} \delta^{\bot}(q_k-q_{n+1}-\sum_{l=k}^{n}k_l)\mathrm{disc}_{s_{k,n+1}}
{\cal A}_{2\rightarrow 2+n}^{\{-\}}=\\
&=2s\Gamma_{J_0A}^{R_1}\frac{e^{\omega(q_1)(y_0-y_1)}}{q_{1\bot}^2}
\Biggl(\prod_{l=2}^{k}\gamma_{R_{l-1}R_l}^{J_{l-1}}\frac{e^{\omega(q_l)(y_{l-1}-y_l)}}
{q_{l\bot}^2}\Biggr)\langle J_kR_k|\Biggl(\prod_{l=k+1}^{n}e^{\widehat{\cal K}
(y_{l-1}-y_l)}\widehat{\cal J}_l\Biggr)e^{\widehat{\cal
K}(y_{n}-y_{n+1})}|J_{n+1}B\rangle. \label{disc-example}
\end{split}
\end{equation*}
Here the ket-states $|J_{n+1} B\rangle$  and the bra-states   $\langle J_kR_k|$
denote   the impact factors for the particle-jet $B\rightarrow J_{n+1}$ and the
Reggeon-jet  $R_k\rightarrow J_k$ transitions respectively,    $\widehat{\cal K}$
and $\widehat{\cal J}_l$  are  the operators of the BFKL kernel and the jet
production. The states are defined by their projections on the two-Reggeon states
with the normalization \eqref{orthonormality} and the operators are specified by
their matrix elements.

The discontinuity  in any  $s_{i,j}$ - channel  $(i<j)$  can be obtained from \eqref{disc-example} by an  appropriate substitution.  If $i=0$, $k$ must be changed on   $i$,  all factors besides $2s$ on the left from $\langle J_kR_k|$ must be omitted and
$\langle J_kR_k|$ must be replaced by $\langle J_0 A|$; if $j< n+1$, $n$ must be changed on   $j-1$ and $|J_{n+1}B\rangle $ must be replaced by
$\frac{e^{\omega(q_{j+1})(y_j-y_{j+1})}}{q_{(j+1)\bot}^2}
\Biggl(\prod_{m=j+2}^{n+1}\gamma_{R_{m-1}R_m}^{J_{m-1}}\frac{e^{\omega(q_m)(y_{m-1}-y_m)}}
{q_{m\bot}^2}\Biggr)\Gamma_{J_{n+1}B}^{R_{n+1}}$.

The BFKL kernel consists of two parts, \be \hat{\cal K}=\omega(\hat r_1)+\omega(\hat
r_2)+\hat{\cal K}_r, \label{virtual + real} \ee where the  ``virtual'' part is given
by the gluon trajectories  and the  ``real'' part ${\cal K}_r$  appears  from real
particle production. In the NLO \be
  \hat{\cal K}_r\ =\ \hat{\cal
K}^{\Delta}_r-\hat{\cal K}^{B}_r\hat{\cal K}^B_r \Delta\,,  \label{subtracted real
kernel} \ee where $\Delta\gg 1$ is an auxiliary parameter serving for separation of
QMRK from pure MRK, ${\cal K}^B_r$ is the LO (Born) real kernel and \be \langle
{\cal G}_1{\cal G}_2|\hat{\cal K}^{\Delta}_r|{\cal G}'_1{\cal G}'_2\rangle
=\delta^{\bot}(r_{1}+r_{2}-r'_{1}-r'_{2}) \sum_J\!\int\!\! \gamma^J_{{\cal G}_1{\cal
G}'_1}\gamma_J^{{\cal G}_2{\cal G}'_2}
\frac{d\phi_J}{2(2\pi)^{D-1}}\theta(\Delta-\Delta_J). \label{real kernel} \ee Here
the  sum is taken over all possible jets and over  all discrete quantum numbers of
these jets,  $\gamma_J^{{\cal G}_2{\cal G}'_2}$ is the effective vertex for
absorption of  the jet $J$ in the Reggeon transition ${\cal G}'_2\rightarrow {\cal
G}_2$ which is related to $\gamma^{\bar J}_{{\cal G}_2{\cal G}'_2}$ by the change of
signs of longitudinal momenta and the corresponding change of wave functions; \be
 d\phi_J=(2\pi)^D\delta^{D}\Bigl(l_J-\sum_i l_i\Bigr)\frac{1}{n!}\frac{dl^2_J}
 {2\pi}\prod_i\frac{d^{D-1}l_i}{2l^0_i(2\pi)^{D-1}}~,   \label{phase space}
\ee
where  $l_i$  are the jet particle momenta, $n$ is a number of identical particles in the jet;    $\Delta_J$  in \eqref{real kernel}  is the interval between the rapidities $z_i=\frac{1}{2}\ln[{l^+_i}/{l^-_i}]$  of the jet particles.  In the Born kernel
the second term in \eqref{subtracted real kernel} is omitted  and
only  one-gluon production in the LO is accounted in \eqref{real kernel}.

Formally the representation of the kernel by Eqs.  \eqref{virtual +
real}--\eqref{phase space} remains the same as in QCD.  The difference is in
appearance of new Reggeon vertices in the sum  over $J$ in \eqref{real kernel} and
in the changes of the gluon trajectory and of the QCD Reggeon vertices because of
dependence of the fermion contributions on representation of the colour group and
appearance of scalar contributions.   The same  applies to the representations  of
the impact factors  and the operator of the jet production.  Remind that they have
to be taken in the NLO in the case  of one-particle  jets and in the LO in the case
of two-particle jets.  The  particle-particle impact-factor for the $B\rightarrow
B^\prime$ ($B$ and $B'$ can be two-particle jets as well) transition  is represented
by the  ket-state $|\bar B^\prime B\rangle$    defined as \be |\bar B^\prime
B\rangle\ =\ |\bar B^\prime B\rangle^{\Delta}-\left({\omega(\hat
r_{1\bot}^2)}\ln\left|\frac{\hat r_{1\bot}}{q_{B\bot}}\right| +{\omega(\hat
r_{2\bot}^2)}\ln\left|\frac{\hat r_{2\bot}}{q_{B\bot}}\right|
 + \hat{\cal K}^B_r\;\Delta\right)|\bar B^\prime
B\rangle^{B}, \label{BB}
\ee
where $|\bar B^\prime
B\rangle^{B}$ is the LO (Born) impact factor and
\be
\langle {\cal G}_1{\cal G}_2|\bar B^\prime B\rangle^{\Delta} =
\delta^{\bot}(q_{B}-r_{1}-r_{2}) \sum_{J}\!\int\!\!\left(\Gamma^{{\cal
G}_1}_{J B} \Gamma^{{\cal G}_2}_{B' J }-\Gamma^{{\cal G}_2}_{J B}
\Gamma^{{\cal G}_1}_{B' J }\right)
 d\phi_{ J}\prod_l\theta\Big(\Delta-(z_l-y_B)\Big)\,.
 \label{BBdelta}
\ee Here $q_{B}=p_{B^\prime}-p_B$, $z_l$ are the rapidities of particles in the
intermediate jets and $y_B=\ln |q_{B\bot}|/(\sqrt 2 p_B^-)$. case when  $B$ or $B'$
is a two-particle jet, only the first term must be kept in Eq.~(\ref{BB}); moreover,
only the Born approximation for this term  must be taken in  Eq.~(\ref{BBdelta}).

For completeness let us present  the impact-factor of the $A\rightarrow A'$
transition, although it is not necessary since it  can obtained from \eqref{BB},
\eqref{BBdelta} by the ``left $\leftrightarrow$ right'' exchange, which means
$|\;\rangle \leftrightarrow \langle\;|,\; A \leftrightarrow B, \;
\x_i\leftrightarrow -\x_i,\; z_l \leftrightarrow -z_l \; y_A \leftrightarrow -y_B,
\; \q_i\leftrightarrow -\q_i, + \leftrightarrow - $.
\begin{equation}
\langle A^\prime \bar A|\ =\ \langle A^\prime \bar A|^\Delta -\langle A^\prime \bar
A|^B\left({\omega(\hat r_{1\bot}^2)} \ln\left|\frac{\hat r_{1\bot}}{q_{A\bot}}\right|
+{\omega^B(\hat r_2)}\ln\left|\frac{\hat r_{2\bot}^2}{q_{A\bot}}\right| + \hat{\cal
K}_r^B\;\Delta\right),
 \label{AA}
\end{equation}
\begin{equation}
\langle A^\prime \bar A|{\cal G}_1{\cal G}_2\rangle^{\Delta} =
\delta^{\bot}(q_{A}-r_{1}-r_{2})\sum_{\tilde A}\!\int\!\!\left( \Gamma^{{\cal
G}_1}_{\tilde AA} \Gamma^{{\cal G}_2}_{A' \tilde A }-\Gamma^{{\cal G}_2}_{\tilde A A }
\Gamma^{{\cal G}_1}_{A' \tilde A }\right)
 d\phi_{ \tilde A}\prod_l\theta\Big(\Delta -(y_A-z_l)\Big),
 \label{AAdelta}
\end{equation}
where $q_{A}=p_A-p_{A^\prime}$, $y_A=\ln(\sqrt 2 p_A^+/|q_{A\bot}|)$.

Accordingly, the Reggeon-particle impact factors are defined as
\[
 |\bar J_i R_{i+1}\rangle \ = \ |\bar J_i R_{i+1}\rangle^\Delta
 -\left(\frac{\omega(q_{(i+1)\bot}^2)}{2}\ln\left|\frac{k^2_{i\bot}}
{|q_{(i+1)\bot}-\hat r_{1\bot}||q_{(i+1)\bot}-\hat r_{2\bot}|}\right| \right.
\]
\begin{equation}
 \left. -\frac{\omega(\hat
r_{1\bot}^2)}{2}\ln\left|\frac{k^2_{i\bot}} {|q_{(i+1)\bot}-\hat r_{1\bot}|\hat
r_{1\bot}}\right|-\frac{\omega(\hat r_{2\bot}^2)}{2}\ln\left|\frac{k^2_{i\bot}}
{|q_{(i+1)\bot}-\hat r_{2\bot}|\hat r_{2\bot}}\right| + \hat{\cal
K}_r^{B}\Delta\right)|\bar J_i R_{i+1}\rangle^B\,, \label{J R}
\end{equation}
\begin{eqnarray}
&& \hspace*{-1cm} \langle {\cal G}_1 {\cal G}_2|\bar J_i R_{i+1}\rangle^\Delta\ =\
\delta^{\bot}(q_{(i+1)}+k_{i}-r_{1}-r_{2})\times
\nonumber\\
&&\times\ \sum_{ J}\int \left(\gamma^{J }_{{\cal G}_1 R_{i+1}} \Gamma^{{\cal G}_2}_{J_i J
}-\gamma^{J }_{{\cal G}_2R_{i+1}} \Gamma^{{\cal G}_1}_{J_i J}\right) d\phi_J
\prod_l\theta\Big(\Delta-(z_l-y_i)\Big),\label{J R delta}
\end{eqnarray}
and
\[
\langle J_i  R_i| \ = \ \langle J_i R_i|^\Delta-\langle J_i
 R_i|^ B\left(\frac{\omega(q_{i\bot}^2)}{2}\ln\left|\frac{k^2_{i\bot}}
{|q_{i\bot}-\hat r_{1\bot}||q_{i\bot}-\hat r_{2\bot}|}\right| \right.
\]
\begin{equation}
 \left. -\frac{\omega(\hat
r_{1\bot}^2)}{2}\ln\left|\frac{k^2_{i\bot}} {|q_{i\bot}-\hat r_{1\bot}|\hat
r_{1\bot}}\right|-\frac{\omega(\hat r_{2\bot}^2)}{2}\ln\left|\frac{k^2_{i\bot}} {|q_{i\bot}-\hat
r_{2\bot}|\hat r_{2\bot}}\right| + \hat{\cal K}_r^{B}\Delta\right), \label{R J}
\end{equation}
\begin{eqnarray}
&& \hspace*{-1cm} \langle J_i  R_i|{\cal G}_{1}{\cal G}_{2}\rangle^{\Delta}\ =\
 \delta^{\bot}(r_{1}+r_{2}-q_{i}+k_{i})\times
\nonumber\\
&& \times\ \sum_{ J}\int \left(\gamma^{J }_{R_i{\cal G}_1} \Gamma^{{\cal G}_2}_{J_i J}
-\gamma^{J }_{R_i{\cal G}_2} \Gamma^{{\cal G}_1}_{J_i J }\right) d\phi_{ J
}\prod_{l}\theta\Big(\Delta -(y_i-z_l)\Big). \label{R J delta}
\end{eqnarray}

And finally, the  operators ${\hat{\cal J}}_i$  for production of jets $J_i$ are
defined as
\begin{eqnarray}
&&\hspace*{-0.5cm} \hat{\cal J}_i = \hat{\cal J}_i^{\Delta}-\left( \hat{\cal
K}^{B}_r\hat{\cal J}^B_i+\hat{\cal J}^{B}_i\hat{\cal K}^{B}_r\right)\Delta, \quad \langle
{\cal G}_{1}{\cal G}_2|\hat{\cal \cal J}^{\Delta}_i|{\cal G}^\prime_{1}{\cal G}^\prime_{2}\rangle =\delta^{\bot}(r_{1}+r_{2}-k_{i}-r^\prime_{1}-
r^\prime_{2})\times
\nonumber\\
&& \times\quad \Bigg[ \gamma^{J_i}_{{\cal G}_1{\cal G}^\prime_1}\delta^{\bot}(r_{2}-r^\prime_{2})
 r_{2\perp}^{~2}\delta_{{\cal
G}_2{\cal G}_2^\prime}+\gamma^{J_i}_{{\cal G}_2{\cal
G}^\prime_2}\delta^{\bot}(r_{1}-r^\prime_{1}) r_{1\perp}^{~2}\delta_{{\cal G}_1{\cal G}_1^\prime}+
\nonumber\\
&&+\quad \sum_{G }\int_{y_i-\Delta}^{y_i+\Delta}
\frac{dz_{G}}{2(2\pi)^{D-1}}\left(\gamma^{\{ J_i G\}}_{{\cal G}_1{\cal
G}^{\prime}_1}\gamma^{{\cal G}_2{\cal G}^\prime_2}_G+ \gamma^G_{{\cal G}_1{\cal
G}^\prime_1}\gamma^{{\cal G}_2{\cal G}^\prime_2}_{J_i G}\right)\Bigg]. \label{jet-delta}
\end{eqnarray}
Here the last term appears only in the case  when $J_i\equiv G_i$ is a single gluon,
the sum in this term goes over quantum numbers of the intermediate gluon $G$ and the
vertices must be taken  in the Born approximation. At that $\gamma^{\{ J_i
G\}}_{{\cal G}_1{\cal G}^{\prime}_1}$ is the vertex for production of the jet
consisting of the gluons $G_i$ and $G$, $\;\;\gamma^{{\cal G}_2{\cal
G}^\prime_2}_{G_i G}$ is the vertex for absorption of gluon $G$ and production of
gluon $G_i$ in the ${\cal G}_2\to{\cal G}'_2$ transition; it can be obtained from
$\gamma_{{\cal G}_2{\cal G}^\prime_2}^{\{G_i G\}}$ by crossing with respect to the
gluon $G$.

\subsection{Bootstrap conditions}

In QCD, it was proved in \cite{Fadin:2006bj}  that an infinite number of the
bootstrap relations  \eqref{bootstrap relations} providing the validity of the
multi-Regge form \eqref{A 2-2+n} in the NLLA   is fulfilled if several  bootstrap
conditions are performed. This statement remains correct for  Yang-Mills  theories
containing  fermions and scalars in arbitrary representations   of the colour group
with any Yukawa-type interaction, because formally all components of the
discontinuities entering in the bootstrap relations \eqref{bootstrap relations}
differ from corresponding  components  in QCD only by appearance of new Reggeon
vertices  and by the changes of the gluon trajectory and of the QCD Reggeon vertices
due to dependence of the fermion contributions on representation of the colour group
and emergence of scalar contributions. Moreover, the bootstrap conditions have the
same form as in QCD. They ere the following.

The particle-jet  impact factors  are  proportional to their  Reggeon vertices: \be
 \langle{A'A}| = {g}\langle R_{\omega}(q_A)|  \Gamma^{R}_{A'A}~, \;\; |{B'B}\rangle = {g}\Gamma^{R}_{B'B}|R_{\omega}(q_B)\rangle~, \;\; \label{IF condition}
\ee where $\Gamma^{R}_{A'A}$ and $\Gamma^{R}_{B'B}$   are  the Reggeon vertices,  $
q_A= p_A-p_{A^\prime}, \;\; q_B= p_{B^\prime}-p_{B}, \;\; $ and
$|R_{\omega}(q)\rangle $ are  the universal (process independent) states.

The states $|R_{\omega}(q)\rangle $  are  the   eigenstate of the kernel $\hat{\cal
K}$ with the eigenvalues $\omega(q)$ \be
 (\hat {\cal K} - \omega(q) )|R_{\omega}(q)\rangle =0, \;\; \langle R_{\omega}(q)|
 (\hat {\cal K} - \omega(q) )=0.   \label{eigenstate condition}
\ee

Moreover, they satisfy the orthonormality  relations
\be \frac{g^2 t}{2(2\pi)^{D-1}}\langle
R'_{\omega}(q')|R_{\omega}(q)\rangle=-\omega(t)\delta^{\bot}(q-q')\delta^{RR'}~.
\label{normalization condition} \ee

The Reggeon-particle impact factors and the jet  production vertices satisfy the
conditions
\[
g q_{i\perp}^ {2}\langle R_{\omega}(q_i)|{\hat{\cal J}_i} + \langle {{\cal J}_i}R_i|
= {g}\gamma^{{\cal J}_i}_{R_iR_{i+1}}\langle R_{\omega}(q_{i+1})|,
\]
\be g q^2_{(i+1)\perp} {\hat{\cal J}_i}|R_{\omega}(q_{i+1})\rangle + |{\cal J}_i
R_{i+1}\rangle = {g}\gamma^{{\cal J}_i}_{R_i R_{i+1}} | R_{\omega}(q_i)\rangle~.
\label{RJ condition} \ee The summation over Reggeon colour index  $R$ in the
right-hand sides of Eqs. \eqref{IF condition} and \eqref{RJ condition} is assumed.

\section{Proof of fulfilment of the bootstrap conditions}

In QCD, the bootstrap conditions \eqref{IF condition}-\eqref{normalization
condition} were formulated    in  \cite{Fadin:1998fv, Braun:1998zj, Braun:1999gt,
Braun:1999uz, Fadin:2000ww, Bartels:2003jq} and their fulfilment was proved in
\cite{Fadin:1998jv, Braun:1998zj,  Fadin:1999de,  Fadin:1999df,  Braun:1999uz,
Fadin:2000qy,  Fadin:2000ww, Fadin:2001fv, Fadin:2002hz, Fadin:2003xs}. The
bootstrap conditions \eqref{RJ condition} were derived in \cite{Bartels:2003jq} and
their fulfilment was proved in \cite{Kozlov:2011zza, Kozlov:2012zza, Kozlov:2012zz}.

To  extend  the proof to Yang-Mills   theories of general form one has to take three
steps. First,  one needs to generalize the proof of the  QCD  bootstrap conditions
to the case  of fermions in arbitrary representation of the colour group. Second,
one has to prove that   contributions of scalars in these conditions don't violate
their fulfilment. And third, one has to prove fulfilment of new bootstrap
conditions.

\subsection{Impact-factors for particle-jet transitions}
We have to separate consideration of  one-particle and two-particle jets.
Corresponding impact factors we will call  {\bf particle $\rightarrow $ particle}
and {\bf particle $\rightarrow $ jet} ones. In the NLLA the first ones must be taken
in the NLO, while for the second ones the Born approximation is sufficient.  Let us
start with particle-particle impact factors.

\subsubsection{Particle $\rightarrow $ particle impact-factors}
In QCD they are the gluon and quark ones.   The first of them was  obtained in
\cite{Fadin:1999de}. The derivation presented there permits to generalize the quark
contribution to this impact factor to any  representation of the colour group. Using
also the results of \cite{Kozlov:2014gaa} for the scalar contribution, we obtain
\[
\langle G^\prime  G|{\cal G}_1{\cal G}_2\rangle = \delta^{\bot}(q-r_1-r_2)g^2
e(p_G)_{\perp\mu}e(p_{G^\prime})^*_{\perp\nu}T^R_{G^\prime  G}T^R_{{\cal G}_1{\cal
G}_2}\Biggl\{ - g_\bot^{\mu \nu}\Biggl[1 - \bar{g}^2\frac
{\Gamma^2(1+\epsilon)}{\epsilon\Gamma(1+2\epsilon)}(-q_\bot^2)^\epsilon \biggl[
{\tilde K_1}
\]
\[
 + \left( \left( \frac{
r_{1\bot}^{\:2}}{
q_{\bot}^{\:2}} \right)^\epsilon +
\left( \frac{ r_{2\bot}^{\:2}}{q_{\bot}^{\:2}} \right)^\epsilon - 1
\right)\biggl( \frac{1}{2\epsilon} +
\psi(1+2\epsilon) - \psi(1+\epsilon)+ \frac{a_1}{2(1+2\epsilon)(3+2\epsilon)}
\biggr)
\]
\[
+ \frac{3}{2\epsilon} + 2\psi(1) - \psi(1+\epsilon) - \psi(1+2\epsilon)
-\frac{(1+\epsilon)^2\;a_1+2\epsilon^2
a_2}{2(1+\epsilon)^2(1+2\epsilon)(3+2\epsilon)}
 \biggr]
\]
\begin{equation}
-\left( g^{\mu\nu}_{\perp} - (D-2)\frac{q^\mu_\perp q^\nu_\perp}{q^2_\perp}
\right)\bar{g}^2\frac
{\Gamma^2(1+\epsilon)}{\Gamma(4+2\epsilon)}(-q_\bot^2)^\epsilon\;
\frac{2a_2}{(1+\epsilon)}\biggr\}, \label{eq:GG-nlo}
\end{equation}
where $q=r_1+r_2$ and
\[
{\tilde K_1} = -\frac{(4\pi)^{2+\epsilon}\Gamma(1+2\epsilon)\epsilon\left(-
q_{\bot}^{\:2} \right)^{-\epsilon}}
{4\Gamma(1-\epsilon)\Gamma^2(1+\epsilon)}\int\frac{d^{D-2}l}{(2\pi)^{D-1}}\ln\left
( \frac{ q_{\bot}^{\:2}}
{ l_{\bot}^{2}} \right)\frac{ q_{\bot}^{\:2}}{( l -  r_1)_{\bot}^2( l +
r_2)_{\bot}^2}
\]
\begin{equation}\label{tilde K 1}
 = \frac{1}{2\epsilon}\left(2- \left( \frac{
 r_{1\bot}^{\:2}}{
 q_{\bot}^{\:2}} \right)^\epsilon -
 \left( \frac{ r_{2\bot}^{\:2}}{q_{\bot}^{\:2}} \right)^\epsilon
\right)+ \frac{\epsilon}{2}\ln\left( \frac{
 r_{1\bot}^{\:2}}{q_{\bot}^{\:2}}\right)\ln\left( \frac{
  r_{2\bot}^{\:2}}{q_{\bot}^{\:2}} \right) -4\epsilon^2 \zeta(3) + {\cal O} (\epsilon^3)
 \;.
\end{equation}
Remind that \be a_1=11+7\epsilon -4(1+\epsilon)\xi_f -{\xi_s}, \;\; a_2=1+\epsilon
-2\xi_f +{\xi_s}~ \ee and  the coefficients $a_1$ and  $a_2$ vanish in $N=4$ SYM in
the dimensional reduction.

Comparing \eqref{eq:GG-nlo} with  gluon-gluon-Reggeon vertex \eqref{eq:GRG-nlo} we
see that  the bootstrap relation \eqref{IF condition} is fulfilled if
$$
\langle R_{\omega}(q)|{\cal G}_1{\cal G}_2\rangle =
\delta^{\bot}(q-r_1-r_2)T^R_{{\cal G}_1{\cal G}_2}\biggl( 1 - \bar{g}^2\frac
{\Gamma^2(1+\epsilon)}{\epsilon\Gamma(1+2\epsilon)}(-q_\bot^2)^\epsilon\biggl[
{\tilde K_1} + \left(\left( \frac{ r_{1\bot}^{\:2}}{q_{\bot}^{\:2}}
\right)^\epsilon+ \left( \frac{ r_{2\bot}^{\:2}}{q_{\bot}^{\:2}} \right)^\epsilon -
1 \right)
$$
$$
\times\biggl\{ \frac{1}{2\epsilon} + \psi(1+2\epsilon) - \psi(1+\epsilon) +
\frac{a_1}{2(1+2\epsilon)(3+2\epsilon)}
\biggr\}
$$
\begin{equation}\label{R omega}
- \frac{1}{2\epsilon} + \psi(1) + \psi(1+\epsilon) - \psi(1-\epsilon) -
\psi(1+2\epsilon) \biggr] \biggr)\;.
\end{equation}

The quark impact factor in QCD  was  obtained   in \cite{Fadin:1999df}. The
calculations presented there  can be easily generalized to any quark representation
of the colour group. Scalars give contributions to the quark impact factors due to
their gauge and  Yukawa-type interactions. The first ones  come from vacuum
polarization diagrams only and are  obtained from corresponding quark contributions
by the  replacement $\xi_f\rightarrow \xi_s/(4(1+\epsilon))$. As for the second
ones, fulfilment of the bootstrap conditions \eqref{IF condition} for them was
proved recently \cite{Kozlov:2014gaa} in the  general form,  for all impact factors,
using the analytic properties of the amplitudes whose imaginary parts are associated
with the impact factors and the vertices in the bootstrap conditions \eqref{IF
condition}.  Using these results, we obtain
\[
\langle Q^\prime_f  Q_i|{\cal G}_1{\cal
G}_2\rangle = \delta^{\bot}(q-r_1-r_2)\delta_{fi}g^2\bar{u}_f(p')t^R_i\frac{\not{n}_2}{2p^+}u_i(p)\;T^R_{{\cal G}_1{\cal G}_2} \Biggl[1 - \bar{g}^2\frac
{\Gamma^2(1+\epsilon)}{\epsilon\Gamma(1+2\epsilon)}(-q_\bot^2)^\epsilon \biggl[ {\tilde K_1}
\]
\[
 + \left( \left( \frac{r_{1\bot}^{\:2}}{q_{\bot}^{\:2}} \right)^\epsilon +\left( \frac{ r_{2\bot}^{\:2}}{q_{\bot}^{\:2}} \right)^\epsilon - 1\right)\biggl( \frac{1}{2\epsilon} +\psi(1+2\epsilon) - \psi(1+\epsilon)+ \frac{a_1}{2(1+2\epsilon)(3+2\epsilon)}\biggr)
\]
\[
 + \frac{1}{2\epsilon} +
2\psi(1) - \psi(1+\epsilon) - \psi(1+2\epsilon)
+\frac{a_1-3(3+2\epsilon)}{2(1+2\epsilon)(3+2\epsilon)}
+ \left(\frac{2C_F^i}{N_c}-1\right)\left(\frac{1}{\epsilon} -\frac{3-2\epsilon}{2(1+2\epsilon)}\right)\biggr]\Biggr]
\]
\begin{equation}
+\delta^{\bot}(q-r_1-r_2)\Gamma_{Q'_fQ_i}^{R(Y)} \;g  T^R_{{\cal G}_1{\cal G}_2}\; . \label{eq:QQ-nlo}
\end{equation}
Fulfilment of the bootstrap condition \eqref{IF condition} for the quark  impact
factor follows from comparison of  this result with \eqref{eq:QRQc} and  \eqref{R
omega}.

To obtain the impact factors for scalar particles we  use the results of
\cite{Kozlov:2014gaa}. In this paper they were calculated in SYM and in the special
scheme which simplifies check of the bootstrap conditions (we call it  bootstrap
scheme). Generalization of the results of \cite{Kozlov:2014gaa}  to  any
representations of the colour group for  quarks  and scalars is carried out in the
same way as for the quark impact factors. Going well to the standard scheme with the
help of the equality
\[
\langle R^{B}_{\omega}(q)|\widehat{\cal U}_q|{\cal G}_1{\cal G}_2\rangle = \delta^{\bot}(q-r_1-r_2)T^R_{{\cal G}_1{\cal G}_2}
\bar{g}^2\frac{\Gamma^2(1+\epsilon)}{\epsilon\Gamma(1+2\epsilon)}(-q_\bot^2)^\epsilon
\]
\[
\times\biggl[ - {\tilde K_1}
+ \left(\left( \frac{ r_{1\bot}^{\:2}}{q_{\bot}^{\:2}} \right)^\epsilon+\left( \frac{ r_{2\bot}^{\:2}}{q_{\bot}^{\:2}} \right)^\epsilon \right)
\biggl( \frac{1}{2\epsilon} - \psi(1) - \psi(1+\epsilon) + \psi(1-\epsilon) +
\psi(1+2\epsilon)\biggr)
\]
\be
-\left( \frac{ r_{1\bot}^{\:2}}{q_{\bot}^{\:2}}\right)^\epsilon\ln\left( \frac{ r_{1\bot}^{\:2}}{q_{\bot}^{\:2}}\right)-\left( \frac{ r_{2\bot}^{\:2}}{q_{\bot}^{\:2}}\right)^\epsilon\ln\left( \frac{ r_{2\bot}^{\:2}}{q_{\bot}^{\:2}}\right)
\biggr],
\ee
where
\begin{equation}
 \begin{split}
 &\langle {\cal G}_1'{\cal G}_2'|\widehat{\cal U}_K|{\cal G}_1{\cal G}_2\rangle =g^2\delta^{\bot}(r_1'+r_2'-r_1-r_2)T^a_{{\cal G}_1'{\cal G}_1}T^a_{{\cal G}_2{\cal G}_2'}\frac{r_{1\bot}^{'2}r_{2\bot}^{'2}}{(2\pi)^{D-1}} \biggl(\frac{r_{1\bot}^{'\alpha}}{r_{1\bot}^{'2}}+ \frac{(r_1-r_1')_{\bot}^{\alpha}}{(r_1-r_1')_{\bot}^2}\biggr)\times\\
 &\times \biggl(\frac{r_{2\bot\alpha}^{'}}{r_{2\bot}^{'2}}+ \frac{(r_2-r_2')_{\bot\alpha}}{(r_2-r_2')_{\bot}^2}\biggr) \ln\Big[\frac{K_{\bot}^2}{(r_1'-r_1)_{\bot}^2}\Big]=\frac{1}{2}\ln\Bigl[
 \frac{K_{\bot}^2}{(r_1-r_1')_{\bot}^2}\Bigr]\langle {\cal G}_1'{\cal
 G}_2'|\widehat{\cal K}_r^B|{\cal G}_1{\cal G}_2\rangle
 \label{eq:oU}
 \end{split}
\end{equation}
and ${\tilde K_1}$ is defined in  \eqref{tilde K 1}, we obtain
\[
\langle S'_{r'}S_r|{\cal G}_1{\cal G}_2\rangle  = \delta^{\bot}(q-r_1-r_2)\delta_{r'r}g^2 ({\cal T}_r^R)_{S'_{r'}S_r}\;T^R_{{\cal G}_1{\cal G}_2} \Biggl[1 - \bar{g}^2\frac
{\Gamma^2(1+\epsilon)}{\epsilon\Gamma(1+2\epsilon)}(-q_\bot^2)^\epsilon \biggl[ {\tilde K_1}
\]
\[
 + \left( \left( \frac{
r_{1\bot}^{\:2}}{
q_{\bot}^{\:2}} \right)^\epsilon +
\left( \frac{ r_{2\bot}^{\:2}}{q_{\bot}^{\:2}} \right)^\epsilon
\right)\biggl( \frac{1}{2\epsilon} +
\psi(1+2\epsilon) - \psi(1+\epsilon)
+ \frac{a_1}{2(1+2\epsilon)(3+2\epsilon)}
\biggr)
\]
\[
+2\psi(1) - 2\psi(2+2\epsilon) +
\left(\frac{2C^{r}_S}{N_c}\right)\left(\frac{1}{\epsilon}
-\frac{2}{(1+2\epsilon)}\right)\biggr]
\]
\begin{equation}
+\delta^{\bot}(q-r_1-r_2)\Gamma_{S'_{r'}S_r}^{R(Y)} \;g  T^R_{{\cal G}_1{\cal G}_2}\; . \label{eq:SS-nlo}
\end{equation}
Fulfilment of the bootstrap condition  \eqref{IF condition} for scalar scattering
follows from comparison of  this result with \eqref{eq:SRSc} and  \eqref{R omega}.

\subsubsection{particle $\rightarrow $ jet impact factors}
For   particle $\rightarrow $ jet transitions   $A\rightarrow A'= \{P_1P_2\}$
the bootstrap condition  \eqref{IF condition} takes the form
\begin{equation}
\langle \{P_1P_2\}A|{\cal G}_1{\cal G}_2\rangle=g\Gamma_{\{P_1P_2\}A}^R\langle R_{\omega}(q)|{\cal G}_1{\cal G}_2\rangle\,,
\label{elasticQMRK}
\end{equation}
where
\begin{align}\label{IFQMRK}
&\langle \{P_1P_2\}A|{\cal G}_1{\cal G}_2\rangle=\delta^{\bot}(k-k_1-k_2-r_1-r_2)\Bigl(\sum_{\{A'\}}\Gamma_{\{P_1P_2\}A'}^{{\cal G}_2}\Gamma_{A'A}^{{\cal G}_1}+\\
&+\sum_{\{P_1'\}}\Gamma_{P_1P'_1}^{{\cal G}_2}\Gamma_{\{P'_1P_2\}A}^{{\cal G}_1}+\sum_{\{P'_2\}}\Gamma_{P_2P'_2}^{{\cal G}_2}\Gamma_{\{P_1P'_2\}A}^{{\cal G}_1}\Bigr)-\{{\cal G}_1\leftrightarrow {\cal G}_2\}\,.\notag
\end{align}
As it was already pointed out we need to  consider  Eqs. \eqref{elasticQMRK} and
\eqref{IFQMRK} in the LO only.  In this approximation fulfilment of the bootstrap
conditions \eqref{elasticQMRK} can be proved without  explicit forms of the impact
factors \cite{Kozlov:2014gaa}. Indeed, using the old-fashioned perturbation theory
we can write
\begin{equation}
\Gamma_{\{CD\}B}^R= \sum_{B'}\frac{
V_{\{CD\}B'}\Gamma_{B'B}^{R}}{2\epsilon_{B'}(\epsilon_{C}+\epsilon_{D}-\epsilon_{B'})}+\sum_{C'}\frac{\Gamma_{CC'}^{R} V_{\{C'D\}B}}{2\epsilon_{B}(\epsilon_{B}-\epsilon_{D}-\epsilon_{C'})}
+\sum_{D'}\frac{\Gamma_{DD'}^{R} V_{\{CD'\}B}}{2\epsilon_{B}(\epsilon_{B}-\epsilon_{D'}-\epsilon_{C})}
\,, \label{old form}
\end{equation}
where  $V_{\{BC\}A}$ is  the vertex of the   $A\rightarrow BC$ transition in which
all particle  momenta are on the mass shell and all particle polarizations are
physical. It is easy to see that in the impact factor \eqref{IFQMRK}  the
contributions resulting  from use for $\Gamma_{\{P_1P_2\}A'}^{{\cal G}_2}$  the last
two terms in the representation \eqref{old form} cancel the contributions resulting
from use  for  $\Gamma_{\{P'_1P_2\}A}^{{\cal G}_1}$  and
$\Gamma_{\{P_1P'_2\}A}^{{\cal G}_1}$ the first term  in \eqref{old form}. Further,
the sum of the contributions coming from use  for $\Gamma_{\{P'_1P_2\}A}^{{\cal
G}_1}$  the third term in \eqref{old form} and for $\Gamma_{\{P_1P'_2\}A}^{{\cal
G}_1}$  the second term  cancel each other with account of the antisymmetrization
$\{{\cal G}_1\leftrightarrow {\cal G}_2\}$. After these cancellation, it's easy to
see that  fulfilment of \eqref{elasticQMRK} follows from  the relation \be
\sum_{B}\Gamma_{BA}^{{\cal G}_1}\Gamma_{CB}^{{\cal G}_2}-\{{\cal G}_1\leftrightarrow
{\cal G}_2\} = g T^R_{{\cal G}_1{\cal G}_2}\Gamma_{CA}^{R}\,, \ee which is the LO
bootstrap condition for the particle-particle impact factors.

\subsection{Bootstrap conditions for  the eigenfunction of the BFKL kernel}
Fulfilment of the bootstrap conditions \eqref{eigenstate condition}  and
\eqref{normalization condition}  were proved in QCD in \cite{Fadin:1998jv,
Braun:1998zj, Braun:1999gt, Fadin:2000qy, Fadin:2002hz}. In fact, the proof can be
applied  to Yang-Mills theories with quarks and scalars in any representations of
the colour group and with any Yukawa-type interactions. First,  the kernel $\hat
{\cal K}$, the eigenstate $|R_{\omega}(q)\rangle $  and the eigenvalue $\omega(q)$
don't depend on the Yukawa-type interactions at all.  For the kernel it follows from
its definition \eqref{virtual + real} -- \eqref{real kernel} and from the explicit
form of the Reggeon production vertices presented in  Sections
\ref{subsubsection:{Production vertex}} and  \ref{subsubsection:{Production
vertices}}; for the  trajectory and for the  eigenstate $|R_\omega\rangle$ it is
seen  from their explicit forms presented in  \eqref{omega as double integral} --
\eqref{a omega} and \eqref{R omega}.  Second, it's seen also from these equations
that the quark contributions  to the trajectory and to the eigenfunction depend on
the quark representation only  through $\xi_f$   and the scalar contributions is
obtained from the quark one by the replacement $\xi_f\rightarrow
\xi_s/(4(1+\epsilon))$.   The same is true for the BFKL kernel in the antisymmetric
adjoint representation of the colour group \cite{Gerasimov:2010zzb, Bartels:2012sw}
which enters into the bootstrap condition \eqref{eigenstate condition}. Therefore
generalization of the proof  of fulfilment of the bootstrap conditions
\eqref{eigenstate condition} and \eqref{normalization condition} presented to
Yang-Mills theories with quarks and scalars   in any representations of the colour
group is trivial.

\subsection{Bootstrap conditions for particle production in the central rapidity region}
In the NLLA, the bootstrap conditions \eqref{RJ condition} has  to be fulfilled both
for the production of a single gluon and for the production  of a two-particle jet.
In the last case it has to be considered in the LO. Let's start with this case.

\subsubsection{Two-particle jet  production}
The jets can be two-gluon, quark-antiquark and  two-scalar ones. Let us denote the
particles in  the jet   $P_1$ and $P_2$.  The impact factor for transition of the
Reggeon $R_1$  into the  jet  has the form
\[
\langle \{P_1P_2\}R_1|{\cal G}_1{\cal G}_2\rangle=\delta^{\bot}(q_1-l_1-l_2-r_1-r_2)\biggl(\sum_{\{P'\}}\Bigl[\gamma_{R_1{\cal
G}_1}^{\{P_1P'\}}\Gamma_{P_2P'}^{{\cal G}_2}+\gamma_{R_1{\cal
G}_1}^{\{P'P_2\}}\Gamma_{P_1P'}^{{\cal G}_2} \Bigr]+
\]
\begin{equation}
+\sum_{\{G'\}}\Gamma_{\{P_1P_2\}G'}^{{\cal
G}_2}\gamma_{R_1{\cal G}_1}^{G'}\biggr)-\{{\cal
G}_1\leftrightarrow {\cal G}_2\}~,  \label{R IF QMRK}
\end{equation}
where  $q_1$ is the Reggeon momentum,  $l_1, \; l_2$  are the particle $P_1$ and
$P_2$ momenta respectively, $r_1$ and $r_2$ are momenta of the Reggeized gluons
${\cal G}_2$ and ${\cal G}_2$.   The  vertices  $\Gamma_{P'P}^{R}$   are  defined in
\eqref{eq:GRG-nlo}--\eqref{eq:SRSc} (remind that here  we need them in the Born
approximation only), $\; \Gamma_{R_1 R_2}^{G}$ is given  by  \eqref{gamma B}, the
vertices $\Gamma_{\{P_1P_2\}G}^{R}$   and  $\gamma_{R_1R_2}^{\{P_1P_2\}}$ are
defined in \eqref{A P1 P2} and \eqref{eq:rrpp}. The matrix
element of the jet production operator  entering in the bootstrap condition
\eqref{RJ condition} can be written as
\[
\langle R_{\omega}(q_1)| \widehat{\cal J}_{P_1P_2}|{\cal
G}_1{\cal
G}_2\rangle\!=\!g \delta^{\bot}(q_1-l_1-l_2-r_1-r_2)\biggl(T^{R_1}_{{\cal
G}_1{\cal G}_2'}\frac{1}{(q_1-r_1)_\bot^2}\gamma_{{\cal G}_2'{\cal G}_2}^{\{P_1P_2\}} +
\]
\begin{equation}
+ T^{R_1}_{{\cal
G}_1'{\cal
G}_2'}\frac{1}{(l_1+r_1)_\bot^2(l_2+r_2)_\bot^2}\gamma^{P_1}_{{\cal G}_1'{\cal G}_1}\!\gamma^{P_2}_{{\cal G}_2'{\cal G}_2}\biggr) -\{{\cal
G}_1\leftrightarrow {\cal G}_2\}
~. \label{ME QMRK}
\end{equation}
The second term here exists only when $P_1$ and $P_2$ are gluons.

The impact factor \eqref{R IF QMRK} and the matrix  element \eqref{ME QMRK} contain
six independent colour structures.  The can be chosen as $\{{\mathbf T}^{a}{\mathbf
T}^{{b}}{\mathbf T}^{{c}}\}_{S_1S_2}$, where ${\mathbf T}^i$ are the colour group
generators for produced particles  and $a, b, c $ are permutations of  ${R_1}, {\cal
G}_1, {\cal G}_2$.  Equating the coefficients at  these structures in the left and
right sides of the bootstrap condition one obtains six equation. However, due to
symmetry of the bootstrap condition with respect to interchange $P_1\leftrightarrow
P_2$ and antisymmetry with  respect to interchange ${\cal G}_1\leftrightarrow {\cal
G}_2$,  only two of  these  equations are independent. The structure ${\mathbf
T}^{R_1}{\mathbf T}^{{\cal G}_1}{\mathbf T}^{{\cal G}_2} $  gives
\[
-B_{P_1P_2}(q_1; l_1, l_2+r_{2\bot})+C_{\bot \mu}(r_1, q_1)A^\mu_{P_1P_2}(q_1-r_{1})-\frac{q_{1\bot}^2}{(q_1-r_1)_\bot^2}B_{P_1P_2}(q_1-r_{1\bot}; l_1, l_2)=
\]
\be -B_{P_1P_2}(q_1; l_1, l_2)~. \label{R_1 G_1 G_2} \ee Here $B_{P_1P_2}$ are
defined in equations~(\ref{eq:Bqq}), (\ref{eq:Bgg}), and (\ref{eq:Bss}). Quantities
$A^\mu_{P_1 P_2}(k)$ are defined for quark-antiquark, two gluons, and two scalars in
Eqs.~ (\ref{eq:g-q bar q}),  (\ref{eq:grgg}), (\ref{eq:grss}) correspondingly.
And lastly,
\be C_{\bot \mu}(r_1, q_1)=
-2\left(q_{1\bot}-\frac{q_{1\bot}^2}{(q_1-r_1)_\bot^2}(q_1-r_1)_\bot\right)_\mu~.
\label{C mu perp} \ee Direct substitution of these expressions shows that the
condition \eqref{R_1 G_1 G_2} holds.

The second  equation can  be obtained using the colour   structure  ${\mathbf
T}^{{\cal G}_1}{\mathbf T}^{R_1}{\mathbf T}^{{\cal G}_2} $. It looks as
\be
\begin{split}
&-B_{P_1P_2}(q_1; l_1+r_{1\bot}, l_2)-B_{P_2P_1}(q_1; l_2+r_{2\bot}, l_1)-C_{\bot
\mu}(r_1, q_1)A^\mu_{P_1P_2}(q_1-r_{1})-\\
& -C_{\bot \mu}(r_2,q_1)A^\mu_{P_2P_1}(q_1-r_{2})
+q_{1\bot}^2\left(\frac{B_{P_1P_2}(q_1-r_{1\bot}; l_1, l_2)}{(q_1-r_1)_\bot^2}+
\frac{B_{P_2P_1}(q_1-r_{1\bot}; l_2, l_1)}{(q_1-r_2)_\bot^2}
\right)-\\
&-\frac{\left(e^{*\mu}_{1\bot}C_{\bot
\mu}(r_1,l_1+r_1)\right)\left(e^{*\mu}_{2\bot}C_{\bot
\mu}(r_2,l_2+r_2)\right)}{(l_1+r_1)_\bot^2(l_2+r_2)_\bot^2} =0 ~. \label{G_1R_1 G_2}
\end{split}
\ee Here, the last term in the left-hand side appears   only in  the case of
two-gluon jet production. Check of fulfilment of \eqref{G_1R_1  G_2} can be
performed by direct substitution of the expressions (\ref{eq:Bqq})--(\ref{eq:Bss}),
and \eqref{A P1 P2}--\eqref{C mu perp}. The check can be simplified by taking the
sum of \eqref{G_1R_1 G_2}, \eqref{R_1 G_1 G_2} and  \eqref{R_1 G_1 G_2} with the
substitution $P_1\leftrightarrow P_2, \; r_1\leftrightarrow r_2$, that gives
\[
-B_{P_1P_2}(q_1; l_1+r_{1\bot}, l_2)-B_{P_2P_1}(q_1; l_2, l_1+r_{1\bot})
-B_{P_1P_2}(q_1; l_1, l_2+r_{2\bot})-B_{P_2P_1}(q_1; l_2+r_{2\bot}, l_1)-
\]
\be
-\frac{\left(e^{*\mu}_{1\bot}C_{\bot \mu}(r_1,l_1+r_1)\right)\left(e^{*\mu}_{2\bot}C_{\bot \mu}(r_2,l_2+r_2)\right)}{(l_1+r_1)_\bot^2(l_2+r_2)_\bot^2} = -B_{P_1P_2}(q_1; l_1, l_2)-B_{P_2P_1}(q_1; l_2, l_1)~.
 \label{combined}
\ee

\subsubsection{Bootstrap conditions for the  Reggeon-gluon impact factor}
In QCD, the bootstrap conditions  \eqref{RJ condition} were proved in Refs.
\cite{Kozlov:2011zza, Kozlov:2012zza}. The proof was generalized  for SYM theories
in \cite{Kozlov:2014gaa}.  Here we extend the proof to Yang-Mills theories with
fermions and scalars in  any representations of the gauge group.

First,we note that in the NLO the Yukawa-type  interaction does not play any role in
the conditions  \eqref{RJ condition}. Then, the basic colour structures can be
chosen in the same way as in QCD:
\begin{equation}\label{eq:color}
\Tr[T^{{\cal G}_2}T^GT^{{\cal G}_1}T^{R_1}]\,,\;\frac{N_c}{2}
T_{R_1{\cal G}_1}^{G'}T_{{\cal
G}_2G}^{G'}\,,\;\frac{N_c}{2}T_{R_1{\cal G}_2}^{G'}T_{{\cal
G}_1G}^{G'}.
\end{equation}
The first structure is symmetric with respect to the replacement ${\cal
G}_1\leftrightarrow {\cal G}_2$. The second and third structures, which are referred
to as the tree structures, are chosen to be identical to those in the Born
impact-factors. Convenience of the choice (\ref{eq:color}) is caused by that  the
virtual corrections appear only at the  tree structures and that the coefficients at
the symmetric structure are  antisymmetric with respect to  the replacement  ${r}_1
\leftrightarrow {r}_2$ of the Reggeon  momenta   because the total antisymmetry of
the components of the bootstrap condition \eqref{RJ condition} (see \eqref{J R}
\eqref{jet-delta}).

As well as in \cite{Kozlov:2012zza, Kozlov:2014gaa},  consideration of the bootstrap
condition \eqref{RJ condition} can be  simplified   by using  of the bootstrap
scheme, where
\begin{equation}\label{U-bootstrap}
\langle GR_1|_* =\langle GR_1|(1-\widehat{\cal U}_k), \;\; \langle R_{\omega}(q)|_*
= \langle R_{\omega}(q)|(1-\widehat{\cal U}_k),\;\;  \widehat{\cal G}_* =
(1+\widehat{\cal U}_k)\widehat{\cal G}(1-\widehat{\cal U}_k),
\end{equation}
where $\widehat{\cal U}_K$ is defined in \eqref{eq:oU},  $k$~ is the momentum of the gluon $G$. Use of this  scheme  permits to avoid the  calculation of  the most complicated  integrals both in the Reggeon-gluon  impact-factor  and in the matrix elements of the gluon production operator. In this scheme the transformed  eigenfunction is calculated  exactly in $D=4+2\epsilon$:
\begin{equation}\label{eq:R_omega}
\langle R_{\omega}(q_1)| {\cal G}_1{\cal G}_2\rangle_*=\langle R_{\omega}(q_1)|(1-\widehat{\cal U}_k)|{\cal G}_1{\cal G}_2\rangle=\delta^{\bot}(q_1-r_1-r_2)T_{{\cal G}_1{\cal G}_2}^{R_1}\Biggl(1-\bar{g}^2R_k(r_1,r_2)\Biggr)\,;
\end{equation}
\begin{equation}\label{eq:Rk}
\begin{split}
 &R_k(r_1,r_2)=\bigl(-(r_1+r_2)_{\bot}^2\bigr)^{\epsilon}\frac{\Gamma^2(1+\epsilon)}{\epsilon\Gamma(1+2\epsilon)}\biggl\{\Bigl[\frac{r_{1\bot}^2}{(r_1+r_2)_{\bot}^2}\Bigr]^{\epsilon}\ln\Big[\frac{(r_1+r_2)_{\bot}^2}{r_{1\bot}^2}\Big]+\Bigl[\frac{r_{2\bot}^2}{(r_1+r_2)_{\bot}^2}\Bigr]^{\epsilon}\\
&\times\ln\Big[\frac{(r_1+r_2)_{\bot}^2}{r_{2\bot}^2}\Big]+\biggl(\Bigl[\frac{r_{1\bot}^2}{(r_1+r_2)_{\bot}^2}\Bigr]^{\epsilon}+\Bigl[\frac{r_{2\bot}^2}{(r_1+r_2)_{\bot}^2}\Bigr]^{\epsilon}-1\biggr)\biggl(\frac{1}{\epsilon}+\ln\Big[\frac{k_{\bot}^2}{(r_1+r_2)_{\bot}^2}\Big]\\
&+\psi(1-\epsilon)-\psi(1)+2\psi(1+2\epsilon)-2\psi(1+\epsilon)
+\frac{a_1}{2(1+2\epsilon)(3+2\epsilon)}\biggr)\biggr\}\,.
\end{split}
\end{equation}
 With ${\cal O}(\epsilon)$ accuracy
\[
R_k(r_1,r_2)=\frac{[-k_{\bot}^2]^{\epsilon}}{\epsilon^2}- \frac{1}{2}\ln^2\Big[\frac{k_{\bot}^2(r_1+r_2)_{\bot}^2}{r_{1\bot}^2r_{2\bot}^2}\Big]+ \ln\Big[\frac{r_{1\bot}^2}{(r_1+r_2)_{\bot}^2}\Big]\ln\Big[\frac{r_{2\bot}^2}{(r_1+r_2)_{\bot}^2}\Big]
\]
\be
+{a_1}\left(\frac{1}{6\epsilon} -\frac{4}{9}\right)~. \label{R k 0}
\ee

Using the results of \cite{Kozlov:2011zza}-\cite{Kozlov:2012zz},  \cite{Kozlov:2014gaa}, we obtain for  the part of the transformed
impact-factor with the tree color structure:
\begin{equation}
\begin{split}
&\langle GR_1|{\cal G}_1 {\cal G}_2 \rangle_{*\text{tree}}={\cal
N}_{\mu}T_{R_1{\cal G}_1}^{G'}T_{{\cal
G}_2G}^{G'}\Biggl\{\left(\frac{(q_1-r_1)_{\bot}^{\mu}}{(q_1-r_1)_{\bot}^2}-\frac{q_{1\bot}^{\mu}}{q_{1\bot}^2}\right)\Biggl[1-\frac{\bar g^2}{2}\biggl( \ln\Bigl[\frac{r_{2\bot}^2}{(q_1-r_1)_{\bot}^2}\Bigr]\ln\Bigl[ \frac{k_{\bot}^2}{r_{2\bot}^2}\Bigr]+\\
& +\ln\Bigl[\frac{q_{1\bot}^2}{r_{1\bot}^2}\Bigr] \ln\Bigl[
\frac{(q_1-r_1)_{\bot}^2}{r_{1\bot}^2}\Bigr]\biggr)\Biggr] +\bar g^2 \Biggl[\frac 12
\left(\frac{k_{\bot}^{\mu}}{k_{\bot}^2}-\frac{q_{1\bot}^{\mu}}{q_{1\bot}^2}\right)\biggl( \ln\Bigl[\frac{(q_1-k)_{\bot}^2}{q_{1\bot}^2}\Bigr]\ln\Bigl[ \frac{(q_1-k)_{\bot}^2}{k_{\bot}^2}\Bigr]-\\
&-\ln\Bigl[\frac{r_{2\bot}^2} {(q_1-r_1)_{\bot}^2}\Bigr]\ln\Bigl[\frac{r_{2\bot}^2}{k_{\bot}^2}\Bigr]\biggr)+\Bigl(q_{1\bot}^{\mu}\frac{(q_1,q_1-k)_{\bot}} {q_{1\bot}^2}-k_{\bot}^{\mu}\frac{(k,q_1-k)_{\bot}}{k_{\bot}^2}\Bigr)I(q_{1\bot},k_{\bot})-\\
&-\Bigl((q_1-r_1)_{\bot}^{\mu}\frac{(q_1-r_1,r_2)_{\bot}} {(q_1-r_1)_{\bot}^2}-k_{\bot}^{\mu}\frac{(r_2,k)_{\bot}} {k_{\bot}^2}\Bigr) I(q_{1\bot}-r_{1\bot},k_{\bot})-\Bigl(q_{1\bot}^{\mu}\frac{(q_1,r_1)_{\bot}} {q_{1\bot}^2}-(q_1-r_1)_{\bot}^{\mu}\\
&\times \frac{(r_1,q_1-r_1)_{\bot}}{(q_1-r_1)_{\bot}^2}\Bigr)I(q_{1\bot},r_{1\bot})-\biggl(\frac{(q_1-r_1)_{\bot}^{\mu}} {(q_1-r_1)_{\bot}^2}-\frac{k_{\bot}^{\mu}}{k_{\bot}^2}\biggr)R_k(r_1,q_1-r_1)-V^{\mu}(q_1-r_1,r_2)+\\
&+\biggl(\frac{q_{1\bot}^{\mu}}{q_{1\bot}^2}- \frac{k_{\bot}^{\mu}}
{k_{\bot}^2}\biggr) R_k(r_1,r_2)+V^{\mu}(q_1,q_1-k)\Biggr]\Biggr \}
-{\cal
N}_{\mu} T_{R_1{\cal G}_2}^{G'}T_{{\cal
G}_1G}^{G'}\Biggl\{r_1\leftrightarrow r_2\Biggr \}\,, \label{IFBresult}
\end{split}
\end{equation}
where \be {\cal N}_{\mu}=\delta^{\bot}(q_1-k-r_1-r_2)\, 2g^2
q^2_{1\perp}e^{*}_{\perp\mu}(k)~, \ee $V^{\mu}(q_1,q_2)$  and $R_k(r_1,r_2)$ are
defined   in \eqref{V mu}  and \eqref{R k 0}  respectively,
\[
I(q_{1\perp},q_{2\perp})= \int^{1}_{0}\dfrac{dx}{(xq_{1}+(1-x)q_{2})^{2}_{\perp}}
\ln\Big[\frac{x
q^{2}_{1\perp}+(1-x)q^{2}_{2\perp}}{x(1-x)(q_1-q_2)^{2}_{\perp}}\Big];
\]
\be
I(q_{1\perp},q_{2\perp})=I(q_{1\perp}, q_{1\perp}-q_{2\perp})=I(q_{2\perp}, q_{2\perp}-q_{1\perp})~.
\label{Idefinition}
\ee
Corresponding part of the matrix element of the gluon production operator
is
\begin{equation}
\begin{split}
&g^2q^2_{1\perp}\langle R_\omega(q_1)|\hat{\cal G}|{\cal G}_1 {\cal G}_2 \rangle_{*\text{tree}}={\cal
N}_{\mu}T_{R_1{\cal G}_1}^{G'}T_{{\cal
G}_2G}^{G'}\Biggl\{\left(\frac{q_{1\bot}^{\mu}}{q_{1\bot}^2}-\frac{(q_1-r_1)_{\bot}^{\mu}}{(q_1-r_1)_{\bot}^2}\right)\Biggl[1-\frac{\bar g^2}{2}\biggl( \ln\Bigl[\frac{r_{2\bot}^2}{(q_1-r_1)_{\bot}^2}\Bigr]\\
&
\times\ln\Bigl[ \frac{k_{\bot}^2}{r_{2\bot}^2}\Bigr]+\ln\Bigl[\frac{q_{1\bot}^2}{r_{1\bot}^2}\Bigr] \ln\Bigl[\frac{(q_1-r_1)_{\bot}^2}{r_{1\bot}^2}\Bigr]\biggr)\Biggr]
+\left(\frac{k_{\bot}^{\mu}}{k_{\bot}^2}-\frac{q_{1\bot}^{\mu}}{q_{1\bot}^2}\right)\Biggl[1-\frac{\bar g^2}{2}
\biggl( \ln\Bigl[\frac{(q_1-k)_{\bot}^2}{q_{1\bot}^2}\Bigr]\ln\Bigl[ \frac{(q_1-k)_{\bot}^2}{k_{\bot}^2}\Bigr]-\\
&-\ln\Bigl[\frac{r_{2\bot}^2} {(q_1-r_1)_{\bot}^2}\Bigr]\ln\Bigl[\frac{r_{2\bot}^2}{k_{\bot}^2}\Bigr]\biggr)\Biggr]-\bar g^2\Biggl[\Bigl(q_{1\bot}^{\mu}\frac{(q_1,q_1-k)_{\bot}} {q_{1\bot}^2}-k_{\bot}^{\mu}\frac{(k,q_1-k)_{\bot}}{k_{\bot}^2}\Bigr)I(q_{1\bot},k_{\bot})-\\
&-\Bigl((q_1-r_1)_{\bot}^{\mu}\frac{(q_1-r_1,r_2)_{\bot}} {(q_1-r_1)_{\bot}^2}-k_{\bot}^{\mu}\frac{(r_2,k)_{\bot}} {k_{\bot}^2}\Bigr) I(q_{1\bot}-r_{1\bot},k_{\bot})-\Bigl(q_{1\bot}^{\mu}\frac{(q_1,r_1)_{\bot}} {q_{1\bot}^2}-(q_1-r_1)_{\bot}^{\mu}\\
&\times \frac{(r_1,q_1-r_1)_{\bot}}{(q_1-r_1)_{\bot}^2}\Bigr)I(q_{1\bot},r_{1\bot})-\biggl(\frac{(q_1-r_1)_{\bot}^{\mu}} {(q_1-r_1)_{\bot}^2}-\frac{k_{\bot}^{\mu}}{k_{\bot}^2}\biggr)R_k(r_1,q_1-r_1)-V^{\mu}(q_1-r_1,r_2)\Biggr]\Biggr \}-\\
&
-{\cal
N}_{\mu} T_{R_1{\cal G}_2}^{G'}T_{{\cal
G}_1G}^{G'}\Biggl\{r_1\leftrightarrow r_2\Biggr \}\,. \label{Gresult}
\end{split}
\end{equation}
The forms (\ref{IFBresult}) and \eqref{Gresult} are   suitable to check the
bootstrap condition \eqref{RJ condition}. It's easy to see using them that
\[
\langle GR_1|{\cal G}_1 {\cal G}_2 \rangle_{*\text{tree}}+g^2q^2_{1\perp}\langle R_\omega(q_1)|\hat{\cal G}|{\cal G}_1 {\cal G}_2 \rangle_{*\text{tree}}={\cal
N}_{\mu}T^{R_2}_{{\cal G}_1{\cal
G}_2}T_{R_1R_2}^{G}\Biggl[\left(\frac{k_{\bot}^{\mu}}{k_{\bot}^2}-\frac{q_{1\bot}^{\mu}}{q_{1\bot}^2}\right)
\]
\be \times\Bigl(1-{\bar g^2} R_k(r_1,r_2)\Bigr)+{\bar g^2} V^{\mu}(q_1,q_1-k)
\Biggr] = {g}\gamma^{G}_{R_1R_{2}}\langle R_{\omega}(q_{2})|{\cal G}_1 {\cal G}_2
\rangle_{*}\,. \ee The last equality follows from  \eqref{eq:RRG_nlo}, \eqref{gamma
B}   and \eqref{eq:R_omega}.

For N=4 SYM in the dimensional reduction scheme (\ref{IFBresult})  gives the result
\cite{Fadin:2014gra}
\begin{equation}
\begin{split}
&\langle GR_1|{\cal G}_1 {\cal G}_2 \rangle_{*\text{tree}}={\cal
N}_{\mu}T_{R_1{\cal G}_1}^{G'}T_{{\cal
G}_2G}^{G'}\Biggl\{\left(\frac{(q_1-r_1)_{\bot}^{\mu}}{(q_1-r_1)_{\bot}^2}-\frac{q_{1\bot}^{\mu}}{q_{1\bot}^2}\right)\Biggl[1-\frac{\bar g^2}{2}\biggl( \ln\Bigl[\frac{(q_1-r_1)_{\bot}^2}{k_{\bot}^2}\Bigr]\ln\Bigl[ \frac{k_{\bot}^2}{r_{2\bot}^2}\Bigr]+\\
&
+\ln\Bigl[\frac{q_{1\bot}^2}{r_{1\bot}^2}\Bigr] \ln\Bigl[\frac{(q_1-r_1)_{\bot}^2q_{1\bot}^2}{(k_{\bot}^2)^2}\Bigr]+4\frac{(-k_{\bot}^2)^{\epsilon}}{\epsilon} -6\zeta(2)\biggr)\Biggr]
+\bar g^2 \Biggl[\frac 12
\left(\frac{k_{\bot}^{\mu}}{k_{\bot}^2}-\frac{q_{1\bot}^{\mu}}{q_{1\bot}^2}\right)\biggl( \ln\Bigl[\frac{(q_1-r_1)_{\bot}^2}{r_{2\bot}^2}\Bigr]\\
&\times \ln\Bigl[ \frac{k_{\bot}^2}{r_{2\bot}^2}\Bigr]+\ln\Bigl[\frac{q_{2\bot}^2} {q_{1\bot}^2}\Bigr]\ln\Bigl[\frac{k_{\bot}^2}{q_{2\bot}^2}\Bigr]\biggr)+\Bigl(q_{1\bot}^{\mu}\frac{(q_1,q_1-k)_{\bot}} {q_{1\bot}^2}-k_{\bot}^{\mu}\frac{(k,q_1-k)_{\bot}}{k_{\bot}^2}\Bigr)I(q_{1\bot},k_{\bot})-\\
&-\Bigl((q_1-r_1)_{\bot}^{\mu}\frac{(q_1-r_1,r_2)_{\bot}} {(q_1-r_1)_{\bot}^2}-k_{\bot}^{\mu}\frac{(r_2,k)_{\bot}} {k_{\bot}^2}\Bigr) I(q_{1\bot}-r_{1\bot},k_{\bot})-\Bigl(q_{1\bot}^{\mu}\frac{(q_1,r_1)_{\bot}} {q_{1\bot}^2}-(q_1-r_1)_{\bot}^{\mu}\\
&\times
\frac{(r_1,q_1-r_1)_{\bot}}{(q_1-r_1)_{\bot}^2}\Bigr)I(q_{1\bot},r_{1\bot})\Biggr]\Biggr
\} -{\cal N}_{\mu} T_{R_1{\cal G}_2}^{G'}T_{{\cal
G}_1G}^{G'}\Biggl\{r_1\leftrightarrow r_2\Biggr \}\,, \label{IFBresultSYM}
\end{split}
\end{equation}
and \eqref{Gresult} becomes
\begin{equation}
\begin{split}
&g^2q^2_{1\perp}\langle R_\omega(q_1)|\hat{\cal G}|{\cal G}_1 {\cal G}_2 \rangle_{*\text{tree}}={\cal
N}_{\mu}T_{R_1{\cal G}_1}^{G'}T_{{\cal
G}_2G}^{G'}\Biggl\{\left(\frac{q_{1\bot}^{\mu}}{q_{1\bot}^2}-\frac{(q_1-r_1)_{\bot}^{\mu}}{(q_1-r_1)_{\bot}^2}\right)\Biggl[1-\frac{\bar g^2}{2}\biggl(4\frac{(-k_{\bot}^2)^{\epsilon}}{\epsilon}+\\
&+\ln\Bigl[\frac{(q_1-r_1)_{\bot}^2}{k_{\bot}^2}\Bigr]\ln\Bigl[
\frac{k_{\bot}^2}{r_{2\bot}^2}\Bigr]+\ln\Bigl[\frac{q_{1\bot}^2}{r_{1\bot}^2}\Bigr]
\ln\Bigl[\frac{(q_1-r_1)_{\bot}^2q_{1\bot}^2}{(k_{\bot}^2)^2}\Bigr]
-6\zeta(2)\biggr)\Biggr]+\left(\frac{k_{\bot}^{\mu}}{k_{\bot}^2}-\frac{q_{1\bot}^{\mu}}{q_{1\bot}^2}\right)\times\\
&\times\Biggl[1-\frac{\bar g^2}{2} \biggl(
\ln\Bigl[\frac{(q_1-r_1)_{\bot}^2}{k_{\bot}^2}\Bigr]\ln\Bigl[
\frac{k_{\bot}^2}{r_{2\bot}^2}\Bigr]+\ln\Bigl[\frac{q_{1\bot}^2}
{q_{2\bot}^2}\Bigr]\ln\Bigl[\frac{q_{1\bot}^2}{k_{\bot}^2}\Bigr] +
\ln\Bigl[\frac{q_{2\bot}^2}
{r_{1\bot}^2}\Bigr]\ln\Bigl[\frac{q_{2\bot}^2r_{1\bot}^2} {(k_{\bot}^2)^2}\Bigr]
 +\\
&+4\frac{(-k_{\bot}^2)^{\epsilon}}{\epsilon} -6\zeta(2) \biggr)\Biggr]-\bar g^2\Biggl[\Bigl(q_{1\bot}^{\mu}\frac{(q_1,q_1-k)_{\bot}} {q_{1\bot}^2}-k_{\bot}^{\mu}\frac{(k,q_1-k)_{\bot}}{k_{\bot}^2}\Bigr)I(q_{1\bot},k_{\bot})-\\
&-\Bigl((q_1-r_1)_{\bot}^{\mu}\frac{(q_1-r_1,r_2)_{\bot}} {(q_1-r_1)_{\bot}^2}-k_{\bot}^{\mu}\frac{(r_2,k)_{\bot}} {k_{\bot}^2}\Bigr) I(q_{1\bot}-r_{1\bot},k_{\bot})-\\
&-\Bigl(q_{1\bot}^{\mu}\frac{(q_1,r_1)_{\bot}} {q_{1\bot}^2}-(q_1-r_1)_{\bot}^{\mu}
\frac{(r_1,q_1-r_1)_{\bot}}{(q_1-r_1)_{\bot}^2}\Bigr)I(q_{1\bot},r_{1\bot})\Biggr]\Biggr
\} -{\cal N}_{\mu} T_{R_1{\cal G}_2}^{G'}T_{{\cal
G}_1G}^{G'}\Biggl\{r_1\leftrightarrow r_2\Biggr \}\,. \label{GresultSYM}
\end{split}
\end{equation}

The part of the Reggeon-gluon impact factor  with the symmetric colour structure
comes  from real gluon  production only.  It has the form  \cite{Kozlov:2012zza}
\begin{equation}
\begin{split}
&\langle GR_1| {\cal G}_1 {\cal G}_2 \rangle_{*\text{sym}}=-{\cal N}_{\mu} \frac{2 {\bar g^2}}{N_c}\Tr[T^{{\cal
G}_2}T^{G}T^{{\cal G}_1}T^{R_1}]\int_0^1dx_1\Biggl\{
\frac{(q_1-r_1)_{\bot}^{\mu}}{(q_1-r_1)_{\bot}^2}\frac{\Gamma^2(1+\epsilon)}{\Gamma(1+2\epsilon)}\times\\
&\times\frac{2(-k_{\bot}^2)^{\epsilon}}{\epsilon x_1^{1-2\epsilon}}+\biggl(\frac{x_1k_{\bot}^{\mu}}{(r_2+x_1k)_{\bot}^2}+\frac{(q_1-r_1)_{\bot}^{\mu}}{(q_1-r_1)_{\bot}^2}\frac{(r_{2\bot}^2-x_1k_{\bot}^2)}{(r_2+x_1k)_{\bot}^2}\biggr)\ln\Big[\frac{(r_1+x_2k)_{\bot}^2(q_1-r_1)_{\bot}^2}{q_{1\bot}^2k_{\bot}^2x_2^2}\Big]-\\
&-\frac{(q_1-r_1)_{\bot}^{\mu}}{(q_1-r_1)_{\bot}^{2}}\frac{1}{x_1}\ln\Big[\frac{(r_1+x_1k)_{\bot}^2(r_2+x_1k)_{\bot}^2(r_2+x_2k)_{\bot}^2}{x_2^2r_{1\bot}^2r_{2\bot}^2(k+r_2)_{\bot}^2}\Big]+\frac{k_{\bot}^{\mu}}{k_{\bot}^2}\frac{1}{x_1}\ln\Big[\frac{(r_1+x_1k)_{\bot}^2}{r_{1\bot}^2}\Big]-\\
&-\frac{q_{1\bot}^{\mu}}{q_{1\bot}^2}\frac{1}{x_1}\ln\Big[\frac{(r_1+x_2k)_{\bot}^2(r_1+x_1k)_{\bot}^2}{(r_1+k)_{\bot}^2r_{1\bot}^2}\Big]
\Biggr\}+{\cal N}_{\mu} \frac{2 {\bar g^2}}{N_c} \Tr[T^{{\cal G}_2}T^{G}T^{{\cal G}_1}T^{R_1}]\int_0^1dx_1\Biggl\{r_1\leftrightarrow r_2\Biggr\}\, \label{IFsymresult-1}
\end{split}
\end{equation}
and equal to $-g^2q^2_{1\perp}\langle R_\omega(q_1)|\hat{\cal G}|{\cal G}_1 {\cal G}_2 \rangle_{*\text{sym}}$, that means fulfilment of the bootstrap conditions \eqref{RJ condition}. Note that the symmetric colour structure is nonplanar and therefore vanishes in the limit of a large number of colours.

\section {Conclusion}

In this paper we have presented  the proof of  the multi-Regge form of multiple
production  amplitudes in the  next-to-leading logarithmic approximation. The proof
is carried out  for   Yang-Mills  theories with  fermions and scalars in arbitrary
representations   of the colour group and  with any Yukawa-type interaction.  It is
based on the bootstrap relations  which follow from compatibility of the multi-Regge
form with the $s$-channel unitarity and connect the discontinuities of the multiple
production  amplitudes in invariant masses of various combinations of produced
particles with amplitude derivatives with respect to rapidities of these particles.
The discontinuities are constructed from several blocks which, in turn, are
expressed in terms of the  gauge boson (gluon) trajectory  and the Reggeon
(Reggeized gluon)  vertices. It turns out that  performing  an infinite number of
these relations  is sufficient to fulfill several bootstrap  conditions imposed on
these building blocks. We have presented  explicit  expressions for the   gluon
trajectory, all the Reggeon vertices and all the  blocks entering into the
discontinuities of the multiple production  amplitudes,  and have demonstrated
fulfilment of the bootstrap conditions.

\subsection*{Acknowledgments}

A.~V. thanks the Dynasty foundation and the President Programm for the financial
support.


\begin{thebibliography}{99}

\bibitem{Fadin:1975cb}
V.~S.~Fadin, E.~A.~Kuraev and L.~N.~Lipatov,
Phys.\ Lett.\  B {\bf 60} (1975) 50.
\bibitem{Kuraev:1976ge}
E.~A.~Kuraev, L.~N.~Lipatov and V.~S.~Fadin,
  Zh. Eksp. Teor. Fiz. \textbf{71} (1976) 840 [Sov. Phys. JETP \textbf{44}
(1976) 443].
\bibitem{Kuraev:1977fs}
E.~A.~Kuraev, L.~N.~Lipatov and V.~S.~Fadin,
Zh.\ Eksp.\ Teor.\ Fiz.\ {\bf 72} (1977) 377 [Sov.\ Phys.\ JETP {\bf 45} (1977) 199].
\bibitem{Balitsky:1978ic}
  Balitsky, I.~I. and Lipatov, L.~N.,
  {\it Sov.\ J.\ Nucl.\ Phys.}\  {\bf 28} (1978) 822
  [{\it Yad.\ Fiz.}\  {\bf 28} (1978) 1597].
\bibitem{Lipatov:1976zz}
  L.~N.~Lipatov,
  Sov.\ J.\ Nucl.\ Phys.\  {\bf 23} (1976) 338
   [Yad.\ Fiz.\  {\bf 23} (1976) 642].
\bibitem{Balitskii:1979}
Ya.Ya.~Balitskii, L.N.~Lipatov, and V.S.~Fadin, in {\it Materials of
IV Winter School of LNPI} (Leningrad, 1979) 109.
\bibitem{Anastasiou:2003kj}
  C.~Anastasiou, Z.~Bern, L.~J.~Dixon and D.~A.~Kosower,
  Phys.\ Rev.\ Lett.\  {\bf 91} (2003) 251602
  [hep-th/0309040].
\bibitem{Bern:2005iz}
Z. Bern, L.~J. Dixon and V.~A. Smirnov,
Phys.Rev. D {\bf 72} (2005) 085001 [arXiv:hep-th/0505205].
\bibitem{Bern:2006ew}
  Z.~Bern, M.~Czakon, L.~J.~Dixon, D.~A.~Kosower and V.~A.~Smirnov,
  Phys.\ Rev.\ D {\bf 75} (2007) 085010
  [hep-th/0610248].
\bibitem{Drummond:2006rz}
  J.~M.~Drummond, J.~Henn, V.~A.~Smirnov and E.~Sokatchev,
  JHEP {\bf 0701} (2007) 064
  [hep-th/0607160].
\bibitem{Bern:2007ct}
  Z.~Bern, J.~J.~M.~Carrasco, H.~Johansson and D.~A.~Kosower,
  Phys.\ Rev.\ D {\bf 76} (2007) 125020
  [arXiv:0705.1864 [hep-th]].
\bibitem{Alday:2007hr}
  L.~F.~Alday and J.~M.~Maldacena,
  JHEP {\bf 0706} (2007) 064
  [arXiv:0705.0303 [hep-th]].
\bibitem{Drummond:2007aua}
  J.~M.~Drummond, G.~P.~Korchemsky and E.~Sokatchev,
  Nucl.\ Phys.\ B {\bf 795} (2008) 385
  [arXiv:0707.0243 [hep-th]].
\bibitem{Drummond:2007cf}
  J.~M.~Drummond, J.~Henn, G.~P.~Korchemsky and E.~Sokatchev,
  Nucl.\ Phys.\ B {\bf 795} (2008) 52
  [arXiv:0709.2368 [hep-th]].
\bibitem{Nguyen:2007ya}
  D.~Nguyen, M.~Spradlin and A.~Volovich,
  Phys.\ Rev.\ D {\bf 77} (2008) 025018
  [arXiv:0709.4665 [hep-th]].
\bibitem{Brandhuber:2007yx}
  A.~Brandhuber, P.~Heslop and G.~Travaglini,
  Nucl.\ Phys.\ B {\bf 794} (2008) 231
  [arXiv:0707.1153 [hep-th]].
\bibitem{Drummond:2007au}
  J.~M.~Drummond, J.~Henn, G.~P.~Korchemsky and E.~Sokatchev,
  Nucl.\ Phys.\ B {\bf 826} (2010) 337
  [arXiv:0712.1223 [hep-th]].
\bibitem{Drummond:2007bm}
  J.~M.~Drummond, J.~Henn, G.~P.~Korchemsky and E.~Sokatchev,
  Phys.\ Lett.\ B {\bf 662} (2008) 456
  [arXiv:0712.4138 [hep-th]].
\bibitem{Drummond:2008aq}
  J.~M.~Drummond, J.~Henn, G.~P.~Korchemsky and E.~Sokatchev,
  Nucl.\ Phys.\ B {\bf 815} (2009) 142
  [arXiv:0803.1466 [hep-th]].
\bibitem{Bartels:2008ce}
J.~Bartels, L.N.~Lipatov and A.~Sabio Vera,
Phys.\ Rev.\ D {\bf 80} (2009) 045002 [arXiv:0802.2065 [hep-th]];
\bibitem{Bartels:2008sc}
 J.~Bartels, L.~N.~Lipatov, A.~Sabio Vera,
Eur.\ Phys.\ J.\  {\bf C65 } (2010) 587 [arXiv:0807.0894 [hep-th]].
\bibitem{Lipatov:2010qg}
L.N.~Lipatov and A.~Prygarin,
Phys.\ Rev.\ D {\bf 83} (2011) 045020 [arXiv:1008.1016 [hep-th]];
\bibitem{Lipatov:2010ad}
  L.~N.~Lipatov and A.~Prygarin,
Phys.\ Rev.\ D {\bf 83} (2011) 125001 [arXiv:1011.2673 [hep-th]].
\bibitem{Bartels:2010tx}
J.~Bartels, L.N.~Lipatov and A.~Prygarin,
Phys.\ Lett.\ B {\bf 705} (2011) 507 [arXiv:1012.3178 [hep-th]].
\bibitem{Fadin:2011we}
V.S.~Fadin and L.N.~Lipatov,
Phys.\ Lett.\ B {\bf 706} (2012) 470 [arXiv:1111.0782 [hep-th]].
\bibitem{Fadin:2013hpa}
  V.~S.~Fadin, R.~Fiore, L.~N.~Lipatov and A.~Papa,
  Nucl.\ Phys.\ B {\bf 874} (2013) 230.
\bibitem{Fadin:2014gra}
  V.~S.~Fadin and R.~Fiore,
  Phys.\ Lett.\ B {\bf 734} (2014) 86
  [arXiv:1402.5260 [hep-th]].
\bibitem{Fadin:2003av}
  V.~S.~Fadin,
  Phys.\ Atom.\ Nucl.\  {\bf 66} (2003) 2017.
\bibitem{Fadin:2006bj}
V.~S.~Fadin, R.~Fiore, M.~G.~Kozlov and A.~V.~Reznichenko,
Phys.\ Lett.\  B {\bf 639} (2006) 74 [arXiv:hep-ph/0602006].
\bibitem{Fadin:2007xy}
  V.~S.~Fadin and R.~Fiore,
  Phys.\ Lett.\ B {\bf 661} (2008) 139
  [arXiv:0712.3901 [hep-ph]].
\bibitem{Gerasimov:2010zzb}
  R.~E.~Gerasimov and V.~S.~Fadin,
  Phys.\ Atom.\ Nucl.\  {\bf 73} (2010) 1214
   [Yad.\ Fiz.\  {\bf 73} (2010) 1254].
\bibitem{Siegel:1979wq}
  W.~Siegel,
  Phys.\ Lett.\ B {\bf 84} (1979) 193.
\bibitem{Fadin:1995dd}
V.~S.~Fadin,
Pisma Zh.\ Eksp.\ Teor.\ Fiz.\  {\bf 61} (1995) 342
[JETP Lett.\  {\bf 61} (1995) 346 ].
\bibitem{Fadin:1995xg}
  V.~S.~Fadin, R.~Fiore and M.~I.~Kotsky,
  Phys.\ Lett.\  B {\bf 359} (1995) 181.
\bibitem{Fadin:1995km}
  V.~S.~Fadin, R.~Fiore and A.~Quartarolo,
  Phys.\ Rev.\  D {\bf 53} (1996) 2729
  [hep-ph/9506432].
\bibitem{Kotsky:1996xm}
M.~I.~Kotsky and V.~S.~Fadin,
  Yad.\ Fiz.\  {\bf 59N6} (1996) 1080
  [Phys.\ Atom.\ Nucl.\  {\bf 59} (1996) 1035].
\bibitem{Fadin:1996tb}
V.~S.~Fadin, R.~Fiore and M.~I.~Kotsky,
  Phys.\ Lett.\  B {\bf 387} (1996) 593
  [hep-ph/9605357].
\bibitem{Blumlein:1998ib}
J. Blumlein,  V. Ravindran and W.~L. van Neerven,
Phys.\ Rev.\   D {\bf 58} (1998) 091502.
\bibitem{DelDuca:2001gu}
V. Del Duca and E.~W.~N. Glover,
JHEP {\bf 0110}   (2001) 035 [hep-ph/0109028].
\bibitem{Fadin:1993wh}
V.~S.~Fadin and L.~N.~Lipatov,
Nucl.\ Phys.\  B {\bf  406}  (1993) 259.
\bibitem{Fadin:1992zt}
V.~S.~Fadin and R.~Fiore,
Phys.\ Lett.\  B {\bf 294}  (1992) 286.
\bibitem{Kozlov:2014gaa}
M.~G.~Kozlov, A.~V.~Reznichenko and V.~S.~Fadin,
Phys.\ Atom.\ Nucl.\  {\bf 77} (2014) 251.
\bibitem{Fadin:1993qb}
  V.~S.~Fadin, R.~Fiore and A.~Quartarolo,
  Phys.\ Rev.\ D {\bf 50} (1994) 2265
  [hep-ph/9310252].
\bibitem{Fadin:2001dc}
  V.~S.~Fadin and R.~Fiore,
  Phys.\ Rev.\ D {\bf 64} (2001) 114012
  [hep-ph/0107010].
\bibitem{Fadin:1996yv}
V.~S. Fadin, R. Fiore and  M. I. Kotsky,
  { Phys.\ Lett.}\  B {\bf  389}  (1996) 737.
\bibitem{DelDuca:1998cx}
V. Del Duca and C.~R. Schmidt,
  {Phys.\ Rev.}\   D {\bf 59}  (1999) 074004.
\bibitem{Fadin:2000yp}
V.~S.~Fadin, R.~Fiore and A.~Papa,
Phys.\ Rev.\  D {\bf  63} (2001) 034001 [hep-ph/0008006].
\bibitem{Fadin:1994fj}
V.~S. Fadin, R. Fiore and A.  Quartarolo,
 {Phys.\ Rev.}\  D {\bf 50} (1994) 5893.
\bibitem{Fadin:1999df}
  V.~S.~Fadin, R.~Fiore, M.~I.~Kotsky and A.~Papa,
  Phys.\ Rev.\ D {\bf 61} (2000) 094006
  [hep-ph/9908265].
\bibitem{Ioffe:2010zz}
  B.~L.~Ioffe, V.~S.~Fadin and L.~N.~Lipatov,
  Cambridge monographs on particle physics, nuclear physics and cosmology, 2010,
  ISBN: 9780521631488.
\bibitem{Fadin:1999de}
  V.~S.~Fadin, R.~Fiore, M.~I.~Kotsky and A.~Papa,
  Phys.\ Rev.\  D {\bf 61} (2000) 094005
  [hep-ph/9908264].
\bibitem{Fadin:1989kf}
  V.~S.~Fadin and L.~N.~Lipatov,
  JETP Lett.\  {\bf 49} (1989) 352
   [Yad.\ Fiz.\  {\bf 50} (1989) 1141]
   [Sov.\ J.\ Nucl.\ Phys.\  {\bf 50} (1989) 712].
\bibitem{Fadin:1996nw}
V.~S.~Fadin and L.~N.~Lipatov,
  Nucl.\ Phys.\  B {\bf 477} (1996) 767
  [hep-ph/9602287].
\bibitem{Fadin:1997hr}
V.~S.~Fadin, R.~Fiore, A.~Flachi and M.~I.~Kotsky,
  Phys.\ Lett.\  B {\bf 422} (1998) 287
  [hep-ph/9711427].
\bibitem{Fadin:1999yv}
V.~S.~Fadin, M.~I.~Kotsky, R.~Fiore and A.~Flachi,
Yad.\ Fiz.\  {\bf 62}   (1999) 1066 [Phys.\ Atom.\ Nucl.\  {\bf 62} (1999) 999].
\bibitem{Fadin:2003xs}
  V.~S.~Fadin, M.~G.~Kozlov and A.~V.~Reznichenko,
  Phys.\ Atom.\ Nucl.\  {\bf 67} (2004) 359
  [hep-ph/0302224].
\bibitem{S:1960}
O. Steinmann,
Helv. Phys. Acta \textbf{33} (1960) 347.
\bibitem{Bartels_1}
J. Bartels,
Phys. Rev. D \textbf{11}  (1975) 2977; 2989.
\bibitem{Bartels_2}
J. Bartels,
Nucl. Phys. B \textbf{151}  (1979) 293; Nucl. Phys. B
\textbf{175} (1980) 365.
\bibitem{Fadin:2002et}
  V.~S.~Fadin,
Proceedings, NATO Advanced Research Workshop, Alushta, Ukraine, August 31-September 6, 2002, ed. R. Fiore {\it et al.}, NATO Science Series, Vol.101, p.235.
\bibitem{Fadin:1998fv}
  V.~S.~Fadin and R.~Fiore,
  Phys.\ Lett.\ B {\bf 440} (1998) 359
  [hep-ph/9807472].
\bibitem{Braun:1998zj}
M.~Braun and G.~P.~Vacca,
  Phys.\ Lett.\  B {\bf 454} (1999) 319
  [hep-ph/9810454].
\bibitem{Braun:1999gt}
M.~A.~Braun,
  hep-ph/9901447.
\bibitem{Braun:1999uz}
M.~Braun and G.~P.~Vacca,
Phys.\ Lett.\  B {\bf 477} (2000)  156[hep-ph/9910432].
\bibitem{Fadin:2000ww}
V.~S.~Fadin, R.~Fiore, M.~I.~Kotsky and A.~Papa,
Phys.\ Lett.\  B {\bf 495} (2000) 329 [hep-ph/0008057].
\bibitem{Fadin:2001fv}
V.~S.~Fadin, R.~Fiore, M.~I.~Kotsky and A.~Papa,
Nucl.\ Phys.\ Proc.\ Suppl.\  {\bf 99A} (2001) 222.

\bibitem{Bartels:2003jq}
  J.~Bartels, V.~S.~Fadin and R.~Fiore,
  Nucl.\ Phys.\ B {\bf 672} (2003) 329
  [hep-ph/0307076].
\bibitem{Fadin:1998jv}
V.~S.~Fadin, R.~Fiore, and A.~Papa,
Phys.\ Rev.\  D {\bf 60} (1999) 074025  [hep-ph/9812456].
\bibitem{Fadin:2000qy}
  V.~S.~Fadin, R.~Fiore and M.~I.~Kotsky,
  Phys.\ Lett.\ B {\bf 494} (2000) 100
  [hep-ph/0007312].
\bibitem{Fadin:2002hz}
V.~S.~Fadin and A.~Papa,
Nucl.\ Phys.\  B {\bf 640} (2002) 309 [hep-ph/0206079].
\bibitem{Kozlov:2011zza}
  M.~G.~Kozlov, A.~V.~Reznichenko and V.~S.~Fadin,
  Phys.\ Atom.\ Nucl.\  {\bf 74} (2011) 758
   [Yad.\ Fiz.\  {\bf 74} (2011) 784].
\bibitem{Kozlov:2012zza}
  M.~G.~Kozlov, A.~V.~Reznichenko and V.~S.~Fadin,
  Phys.\ Atom.\ Nucl.\  {\bf 75} (2012) 493.
\bibitem{Kozlov:2012zz}
  M.~G.~Kozlov, A.~V.~Reznichenko and V.~S.~Fadin,
  Phys.\ Atom.\ Nucl.\  {\bf 75} (2012) 850.
\bibitem{Bartels:2012sw}
  J.~Bartels, V.~S.~Fadin, L.~N.~Lipatov and G.~P.~Vacca,
  Nucl.\ Phys.\ B {\bf 867} (2013) 827
  [arXiv:1210.0797 [hep-ph]].
\bibitem{FG:2000}
V.~S. Fadin, D.~A. Gorbachev,
Phys. Atom. Nucl. \textbf{63} (2000) 2157 [Yad. Fiz.
\textbf{63} (2000) 2253].


\end{thebibliography}
\end{document}